\documentclass[preprint]{aastex}

\usepackage{rotating}
\input epsf

\shorttitle{Earth, Moon, Sun, and CV Accretion Disks}
\shortauthors{Montgomery}

\begin{document}

\title{Earth, Moon, Sun, and CV Accretion Disks}
\author{M.M. Montgomery\altaffilmark{1}}
\affil{Department of Physics, Univeristy of Central Florida, Orlando, FL  32816, USA}

\begin{abstract}
Net tidal torque by the secondary on a misaligned accretion disk, like the net tidal torque by the Moon and the Sun on the equatorial bulge of the spinning and tilted Earth, is suggested by others to be a source to retrograde precession in non-magnetic, accreting Cataclysmic Variable (CV) Dwarf Novae systems that show negative superhumps in their light curves.  We investigate this idea in this work.  

We generate a generic theoretical expression for retrograde precession in spinning disks that are misaligned with the orbital plane.  Our generic theoretical expression matches that which describes the retrograde precession of Earths' equinoxes.  By making appropriate assumptions, we reduce our generic theoretical expression to those generated by others, or to those used by others, to describe retrograde precession in protostellar, protoplanetary, X-ray binary, non-magnetic CV DN, quasar and black hole systems.  

We find that spinning, tilted CV DN systems cannot be described by a precessing ring or by a precessing rigid disk.  We find that differential rotation and effects on the disk by the accretion stream must be addressed.  Our analysis indicates that the best description of a retrogradely precessing spinning, tilted, CV DN accretion disk is a differentially rotating, tilted disk with an attached rotating, tilted ring located near the innermost disk annuli.  In agreement with observations and numerical simulations by others, we find that our numerically simulated CV DN accretion disks retrogradely precess as a unit.  Our final, reduced expression for retrograde precession agrees well with our numerical simulation results and with selective observational systems that seem to have main sequence secondaries.  

Our results suggest that a major source to retrograde precession is tidal torques like that by the Moon and the Sun on the Earth.  In addition, these tidal torques should be common to a variety of systems where one member is spinning and tilted, regardless if accretion disks are present or not.  Our results suggest that the accretion disk's geometric shape directly affects the disk's precession rate.

\end{abstract}


\section{Introduction}
Variability in non-magnetic, Cataclysmic Variable (CV) close binaries has been largely attributed to instabilities.  Outbursts in recurrent and classical novae can be caused by thermonuclear runaway on the surface of a white dwarf or by a thermal and viscous instability in the disk.  The latter outburst involves cycling from low-to-high thermal and viscous states (see e.g., Meyer \& Meyer-Hofmeister 1981 and references within), and thus cycling from low-to-high mass transfer rates (see e.g., Lasota 2001), that results in a release of accretion energy as mass is drained onto the white dwarf.  This type of outburst is similar to that experienced by CV Dwarf Novae (DN), but larger in scale, and by transient X-ray binaries that have neutron stars or black holes as the primary star in the binary.  Another type of instability, a tidal instability, results in CV DN SU UMa and Nova-Like (NL) outbursts, of which the larger amplitude and longer duration superoutbursts involve both a thermal-viscous instability and a tidal instability.  Both CV DN SU UMa and NL outbursts and superoutbursts show characterisitc hump-shaped modulations, known as positive superhumps, in their light curves.  Positive superhumps have periods that are a few percent longer than the orbital period.   

In addition to positive superhumps, some non-magnetic CV DN systems (e.g., TT Ari) show negative superhumps in their light curves.  These negative superhumps have a period that is a few percent shorter than the orbital period.   Negative superhumps are suggested to be a consequence of a partially tilted disk (Patterson et al. 1993) or are related to a warped accretion disk (Petterson 1977; Murray \& Armitage 1998; Terquem \& Papaloizou 2000; Murray et al. 2002; Foulkes, Haswell, \& Murray 2006).  Our previous numerical simulations (Wood, Montgomery, \& Simpson 2000; Montgomery 2004, Montgomery 2009) show that negative superhumps can be produced by a disk fully tilted out of the orbital plane.  In Montgomery (2009), we suggest that the source to the negative superhump is extra light from inner annuli of a tilted accretion disk, and this extra light waxes and wanes with the amount of gas stream overflow and particle migration to inner annuli as the secondary orbits.  

Accretion disks have been postulated to warp or tilt via a potpourri of sources.  For example, disk tilt in X-ray binaries can be from gas streaming at an upward angle from the inner Lagrange point for one half of the orbit and at a downward angle for the second half of the orbit (Boynton et al. 1980).  In some CVs, a disk tilt can be held constant by a gas stream that is fed via the magnetic field of the secondary (Barrett, O'Donoghue, \& Warner 1988).  Also for CV systems, a disk tilt instability can result from a coupling of an eccentric instability to Lindblad resonances (Lubow 1992).  A vertical resonant oscillation of the disk midplane can be caused by tidal interactions between a massive secondary and a coplanar primary (Lubow \& Pringle 1993).  A warping instability can be caused by irradiation from the primary (Pringle 1996, 1997).  A warping can be caused by direct tidal forces from a secondary orbiting on an inclined orbit (e.g., Papaloizou \& Terquem 1995, Larwood et al. 1996, Larwood 1997, Larwood \& Papaloizou 1997).  A disk tilt can be induced if the secondary has a body axis that is tilted relative to the orbital plane (Roberts 1974).  A disk warp can be caused by misalignments of the spin axis of a compact and/or magnetized primary and the disk axis (see e.g., Kumar 1986, 1989).  To date, no consensus has been found regarding the source to a disk tilt or warp.  We do acknowledge that Murray \& Armitage (1998) find that CV DN accretion disks do not tilt significantly out of the orbital plane by instabilities.  

Regardless of how an accretion disk tilts or warps, non-coplaner accretion disks should precess in the retrograde direction, that is, the direction opposite to spin and orbital motion.   Precession is the motion of a body axis around a fixed axis in space.  Retrograde precessional motion traces out a cone around the cone's symmetric axis, a fixed axis in space.  Two sources are known to generate retrograde precession.  One is torques acting on the body.  A second kind is fast precession which applies to bodies that are free.  An example of the latter is a coin thrown into the air in such a way that the coin spins fast, but the normal to the coin wobbles.  Of the two, many typically cite torques as the source to retrograde precession.  For example, retrograde precessional activity has been suggested for X-ray binaries (e.g., Katz 1973, Larwood 1998, Wijers \& Pringle 1999), for common systems that produce jets (e.g., Livio 1999), for protostars (e.g., Papaloizou \& Terquem 1995, Larwood et. al. 1996, Larwood 1997), for quasars and supermassive black holes (e.g., Romero et al. 2000, Caproni \& Abraham 2002), for CVs (Barrett et al. 1988, Patterson et al. 1993, Harvey et al. 1995), to name a few.  Katz (1973) suggests that the permanently tilted edge of a disk precesses due to the tidal field of the secondary.  Roberts (1974) suggests that the secondary precesses due to the tidal field of the primary and the tilted disk retrogradely precesses as a consequence.  Barrett et al. (1988) and Kondo et al. (1983) suggest that particles may be orbiting a tilted disk and as a result, the disk retrograde precesses.   Patterson et al. (1993) notice a near 2:1 ratio of the negative-to-positive precessional periods of several CV systems that exhibit both modulations in their light curves, and they note that this ratio is curiously similar to that of the Earth which has a retrograde nodding precessional period $P_{r}$=18.61 yr and a prograde precessional period $P_{p}$=8.85 yr (see e.g., Stacy 1977).  Patterson et al. (1993) suggest that the source to retrograde precession in these CV DN systems may be the same as that which causes the Earth to retrogradely precess.  Petterson (1977) suggests that the Earth's retrograde precession is caused by outside forces, namely tidal torques on the Earth's bulge by the Moon and Sun with the Moon having the most effect due to its closer proximity to Earth.   Hence, Patterson et al. (1993) suggests that a tidal torque by the secondary on a tilted accretion disk could be the source to retrograde precession in CV DN non-magnetic, accreting disks.

Others have found that only portions of their disk retrogradely precesses.  For semidetached binaries, Bisikalo et al. (2004) show that a one-armed spiral shock wave can be generated in inner annuli of cooler disks (i.e., $T \sim 10^{4} K$).  They find that their inner annuli spiral wave precesses retrogradely, but the disk does not.  Based on their numerical simulations, they show that the wave is generated by sheared elliptical orbits.  They suggest that the sheared elliptical orbits are due to retrograde precession within the disk, and the retrograde precession is due to an influence by the mass-losing star by the processes discussed in Kumar (1986) and Warner (1995, 2003).  However, this suggestion cannot be correct because Kumar (1986) finds that retrograde precession is due to the tidal influence of the secondary on a tilted disk's outer annulus, not inner annulus, and because the simulations by Bisikalo et al. (2004) do not  have warped or tilted disks (private communication).   

In this work, we test the idea of Patterson et al. (1993) on the tilted accretion.  We assume the disk is fully tilted out of the orbital plane because our previous works (Wood, Montgomery, \& Simpson 2000; Montgomery 2004, Montgomery 2009) are successful in showing that fully tilted disks retrogradely precess.  In addition, our works also generate negative superhump modulations in our artificial light curves that have the correct shape and negative superhump periods of the right percentage as compared with observations.  In this work we do not explain how accretion disks tilt but instead check to see if the source to retrograde precession is like that which causes the spinning, tilted Earth to retrogradely precess.  In \S 2 of this paper, we develop a generic  theoretical expression to describe retrograde precession using the Earth-Moon-Sun system as a model.  In \S 3, we reduce the generic theoretical expression by considering the geometry of the system and by making appropriate assumptions, thereby obtaining expressions generated or used by others to describe retrograde precession in protoplanetary, protostellar, X-ray, black hole, and non-magnetic accreting CV binary systems.  In \S 4, we summarize our numerical simulations.  In \S 5, we list observational data.  Is \S 6, we compare retrograde precessional values obtained from our theoretical expressions with those obtained from our numerical simulations and with those obtained from observational data.  In \S 7, we provide a discussion and in \S 8, we summarize, conclude, and list future work.

\section[]{Theoretical Expressions for Retrograde Precessing}
\subsection{Binary Systems}
We consider two systems, one being the Earth-Moon-Sun system and one being a non-magnetic, accreting CV DN system where the accretion disk is misaligned with the orbital plane.  In the Earth-Moon-Sun system, the spinning Earth is known to have an oblate spheroidal shape.  The resultant equatorial bulge is not symmetric about Earth's rotation axis.  As such, a net torque on this equatorial bulge results in the rotational axis of the Earth precessing in the opposite sense to the Earth's spin and Earth's orbital directions.   

To simplify the many-body problem of the Earth-Moon-Sun system and to generate like coordinates among the two systems, we only consider net torque effects by the Sun on the Earth.   Figure 1, top panel, exaggerates the Earth and Sun size and distance scales to show details.  In this panel, the Sun is the small sphere on the $\bf{y}$ axis located at a distance $d$ from the origin.  The Earth is the large oblate spheroid that is centered at the origin in the ($\bf{x,y,z}$) coordinate system of the ecliptic.  The $\bf{z}$ axis defines the normal to the ecliptic and orbital motion $\bf{\dot{\omega}}$ is prograde.  The ($\bf{x',y',z'}$) coordinate system ($\bf{y'}$ not shown) is the equatorial system of the Earth.  Earth rotates around the $\bf{z'}$ axis with angular velocity $\bf{\dot{\epsilon}}$.  As shown,  the angle from the $\bf{z}$ to the $\bf{z'}$ axes is the obliquity $\theta$.  Note that $\bf{x}$ and $\bf{x'}$ share the same axes and thus they coincide with the line of equinoxes at this instant in time in the Earth-Sun system (or with the line of nodes in the non-magnetic, accreting CV DN system).   Also note that $\bf{z'}$, $\bf{z}$, and $\bf{y}$ are in the same plane and thus Eulerean rotation is about the $\bf{x,x'}$ axis.  Because the gravitational attraction forces differ from a point on Earth nearer to the Sun, denoted by the force vector $\bf{F_{n}}$, than a point on Earth further from the Sun, denoted by the force vector $\bf{F_{f}}$, torques around the origin result.  

Figure 1, bottom panel, represents the CV DN system and is similar to Figure 1a except that the oblate spheroidal Earth is replaced with a massive primary star $M_{1}$ that has angular velocity $\dot{\epsilon}$ and that is surrounded by a disk of mass $M_{d}$ where $M_{d} \ll M_{1}$.  Also, the Sun is replaced with a secondary star of mass $M_{2}$.  In this non-magnetic CV DN system, the massive primary cannot be seen as it is smaller in size compared to the disk size.  We assume that the separation $d$ of the system is much larger than the thickness of the disk, and we assume that the disk has negligible mass compared to that of the primary.  Like the top panel, this bottom panel is not to scale.

Ultimately we seek the retrograde precession of the line of nodes in the spinning, tilted CV DN system.  To find this precession, we use the precession of the equinoxes in the Earth-Sun system as a model.

\subsection{Angular Momentum and Moment of Inertia Tensor}
For continuous bodies, its angular momentum vector $\bf{L}$ is related to its angular velocity vector $ \bf{\dot{{\Omega}}} $ in any coordinate system through a linear transformation matrix known as the moment of inertia tensor $\bf{I}$, 
\begin{equation}
\bf{L} = \bf{I}\bf{\dot{\Omega}}
\end{equation}
\noindent
Using a right-hand coordinate system whose general coordinates are (x',y',z'), we can identify the diagonal elements of the matrix as the moment of inertia coefficients and the off-diagonal elements as the products of inertia.  If we assume, for example, that $\bf{\dot{\Omega}}$ is about the $\bf{x'}$ axis and therefore $\bf{L}$ is along the $\bf{x'}$ axis, then
\begin{equation}
L_{x'} = I_{x'x'}\dot{\Omega}_{x'} + I_{x'y'}\dot{\Omega}_{y'}+I_{x'z'}\dot{\Omega}_{z'}.
\end{equation}

If the density $\rho$ of a solid is known, then the moment of inertia can be found from \( I=\int \rho_{r'} r_{\perp}^{'2} dV' \) where $r'_{\perp}$ is the perpendicular radial distance from any rotational axis, $dV'$ is small volume element, and density is a function of radius.  In cartesian coordinates, the moment of inertia tensor is 
\begin{equation}
I=\int \rho(x',y',z')
\left[ 
\begin{array}{ccc}
  	y'^{2}+z'^{2} & -x'y' & -x'z' \\
	-x'y' & z'^{2}+x'^{2} & -y'z' \\
	-x'z' & -y'z' & x'^{2}+y'^{2}
         \end{array} \right] dV'
\end{equation}
\noindent
where dV'=dx'dy'dz'.  If cylindrical coordinates are desired, then we can integrate from $(-h_{c}/2)$ to $(h_{c}/2)$ along the $z'$ axis where $h_{c}$ is the height of the cylinder, from $-r_{d}$ to $r_{d}$ along the $y'$ axis where $r_{d}$ is the radius of the cylinder, and from $-\sqrt{r_{d}^{2}-y'^{2}}$ to $\sqrt{r_{d}^{2}-y'^{2}}$ along the $x'$ axis.   However, the solution to this integral is most easily found using curvilinear coordinates $q_{1}=r'$, $q_{2}=\phi$, $q_{3}=z'$ with scale factors $h_{1}=1$, $h_{2}=r'$, and $h_{3}=1$ for a circular cylinder.  In this solution, $dV'=r' dr' d\phi dz'$, $x'=r'\cos \phi$, $y'=r'\sin \phi$, and $z'=z'$.  The limits of integration are $-h_{c}/2 \rightarrow h_{c}/2$ on $dz'$, $0 \rightarrow 2\pi$ on $d\phi$, and $0 \rightarrow r_{d}$ on $dr'$.  

\subsection{Potential}
In the Earth-Sun system, the potential we seek is that of the Earth and thus we center this attracting body at the origin of our coordinate system.  As shown in Figure 2, the attracting body has mass $m$; an elemental mass $dm$ is located at a distance $r'$ from the center of mass of the attracting body and at an angle $\alpha$ to the line of centers connecting the center of mass of the attracting body to the center of mass of a far-away point mass object $M$.  We assume in the Earth-Sun system that the Sun is located at a distance $d$ where $d>>r'$.  The potential $V$ at the far away point mass object is

\begin{equation}
V = - \frac{GMm}{d} -\frac{GM}{2d^{3}}(I_{x'x'}+I_{y'y'}+I_{z'z'}-3I)
\end{equation}

\noindent
where we have assumed that Earth has uniform density $\rho$ and rotation is possible about the line of centers.  In this equation, $G$ is the universal gravitational constant.  

This potential is that of MacCullagh, a result that is applicable for slightly non-spherical celestial bodies and for $d>>r'$.  This potential is valid for the oblate spheroidal Earth in the Earth-Moon-Sun system.  We assume that this potential is also valid for the spinning, tilted CV DN system even though the primary is surrounded by a disk (and hence the geometry is by far not spherical) and the radius of the white dwarf primary plus accretion disk is not significantly less than the distance to the point mass secondary star.  As the first term in the potential dominates over the other terms (of which only the second term is shown as all others are assumed negligible) and as $M_{d} \ll M_{1}$, the potential is most affected by $M$ and $ m=M_{1}+M_{d} \sim M_{1}$.  As $M_{1}$ is assumed spherical and of uniform density and as the secondary is significantly further away than compared with the radius of the primary, we accept this potential as valid for the spinning, tilted CV DN system.  We note that the second term in Equation (4) is not negligible as the disk's geometric shape affects the potential via the moment of inertia.  We also note that as the radius of the disk is dependent on the mass ratio of the system, with smaller mass ratios yielding larger disk radii, accretion disks in small mass ratio systems may be subject to additional perturbations.  These additional correction terms are not considered in this work.  

\subsection{Net Torque}
In Figure 2, the Earth in the Earth-Sun system and the primary in the CV DN system are the attracting mass bodies that are centered at the origin.  The Sun and the secondary star are considered to be the far away point mass objects.  As the potential is evaluated at the far away point mass object, the force, which is proportional to the gradient of the potential, is on the far away point mass object.  By Newton's Third Law, the far away point mass object exerts an equal and opposite force on the attracting body.  As the forces are not co-linear, a torque about the origin arises.  Figure 1 is exaggerated to show non-colinear forces.  

Because the spinning Earth has rotational flattening and because the Earth's spin axis is tilted, the force pulling on the near-side of Earth (i.e., the side closer to the Sun) is greater than the force pulling on the far-side of the Earth, and a net non-zero torque results about the $\bf{x}$ axis (see Figure 1, top panel).  In the spinning, tilted CV DN system, a torque about the origin arises for the same reasons - because the disk is tilted and spinning and because this combination results in non-colinear near and far side forces on the disk (see Figure 1, bottom panel).  As shown in Figure 1, the body is rotating counter-clockwise about the $\bf{z'}$ body axis, and the body moment of inertia coefficients are along the body axes ($\bf{x', y', z'}$).  

By definition, the net torque around the origin of the body is \( \bf{\Gamma} = \bf{r'}\)  x  \( \bf{F} \) where any force $\bf{F} =  - \bf{\nabla} ($$V$).  As $\bf{r'}$ x $\bf{r'}$ = 0, then the $I$ in Equation (4) is the only term that is not just a function of radius but also a function of (x',y',z').  Therefore, we reduce Equation (4) to
\begin{eqnarray}
V & = &\frac{3GMI}{2d^{3}} \\
   &  = &\frac{3GM}{2d^{5}}(I_{x'x'}x'^{2} + I_{y'y'}y'^{2} + I_{z'z'}z'^{2})
\end{eqnarray}
\noindent
where we have evaluated the potential at a distance $r'=d$ from the origin.  The net torque in the ($\bf{x',y',z'}$) coordinate system is
\begin{eqnarray}
\bf{\Gamma} & = & \bf{\Gamma_{x'}} + \bf{\Gamma_{y'}} + \bf{\Gamma_{z'}} \\
                         & = &   \frac{-3GM}{d^{5}} (I_{z'z'}-I_{y'y'})(y'z') \bf{x'} - \nonumber \\
                         &     & \frac{3GM}{d^{5}} (I_{x'x'}-I_{z'z'})(x'z') \bf{y'} - \nonumber \\
                         &     & \frac{3GM}{d^{5}} (I_{y'y'}-I_{x'x'})(x'y') \bf{z'} .
\end{eqnarray}
\noindent
By employing transformation equations (see e.g., Goldstein 1980), we can rotate the net torque from the ($\bf{x',y',z'}$) coordinate system to that of the ($\bf{x,y,z}$) coordinate system (see Figure 1).  In this transformation, we assume that the obliquity angle $\theta$ remains constant, the orbit is circular, and the mean orbital motion is $\dot{\omega}$ as shown in Figure 1.  The resultant net torque in the ($\bf{x,y,z}$) coordinate system becomes  \( \bf{\Gamma} =  \bf{\Gamma_{x}} + \bf{\Gamma_{y}} + \bf{\Gamma_{z}}  \) with component magnitudes 
\begin{eqnarray}
\Gamma_{x} & = &   \frac{3GM}{d^{3}} (I_{z'z'}-I_{y'y'})[\sin\theta \cos\theta \sin^{2}(\dot{\omega} t + \psi)]  \\
\Gamma_{y}  &  =  &   \frac{3GM}{d^{3}} (I_{x'x'}-I_{z'z'})[\cos(\dot{\omega} t + \psi)\sin(\dot{\omega} t + \psi) \sin\theta]  \\
\Gamma_{z}  &   =  &  \frac{-3GM}{d^{3}} (I_{y'y'}-I_{x'x'})[\cos(\dot{\omega} t + \psi)\sin(\dot{\omega} t + \psi) \cos\theta] .
\end{eqnarray}
\noindent
where $\psi$ is some phase angle, and $t$ is time, and $\omega$ is an angle that increases in the orbital direction from perihelion.  In these equations, we have neglected any variations over one orbit because the precessions we seek are over long periods of time, much longer than one orbital period.  

To simplify the equations further, we assume symmetry about the $\bf{z'}$ rotation axis, and we assume a circular disk.  Therefore, $I_{x'x'}=I_{y'y'}$ and $\Gamma_{z}$ is zero.  We cannot simplify $\Gamma_{y}$ any further but we we can substitute the identity \( \sin(2\theta) = 2\sin\theta\cos\theta \) into $\Gamma_{x}$.  Because the magnitude of this torque is not constant, we need to take averages.  For example, the magnitude is a minimum at either of the equinoxes and a maximum at either of the solstices in the Earth-Sun system.  Likewise, the magnitude is a minimum when the primary, secondary, and the disk's line of nodes align and a maximum when the primary, secondary, and disk's line of antinodes align.  Therefore, we can find the average of $\Gamma_{x}$ and $\Gamma_{y}$ over a time period of one orbit using an integral average of some continuous function f(t) over an interval $\tau$, \( <f(t)> = \frac{1}{\tau} \int_{0}^{\tau} f(t) dt \).  If we assume in $\Gamma_{x}$ that only the $\sin^{2}(\dot{\omega} t + \psi)$ is not constant and we assume the phase constant is zero, then we find that \( < \sin^{2}(\dot{\omega} t + \psi) > =  1/2 \).  For the same time interval, we find that $\bar{\Gamma}_{y} =0$.  Of Equations (9)-(11), only $\Gamma_{x}$ is left,
\begin{equation}
\bar{\Gamma}_{x} =  \frac{3GM}{4d^{3}}(I_{z'z'}-I_{x'x'})\sin(2\theta).
\end{equation}
\noindent
We can use unit vector notation and employ the cross product definition,
\begin{eqnarray}
\hat{\bf{z'}} \mathrm{x} \hat{\bf{z}} &= & (z'z \sin\theta) \hat{\bf{x'}}   \nonumber \\
 & = & (\sin\theta) \hat{\bf{x'}},
\end{eqnarray}
\noindent
to obtain the component of the net torque along $\bf{\hat{x}}$,
\begin{equation}
\bar{\Gamma}_{x} =  \left[ \frac{3GM}{2d^{3}}(I_{z'z'}-I_{x'x'})\cos(\theta) \right].
\end{equation}
\noindent
For the Earth-Sun system, $M=M_{sun}$ and is not in solar masses.   For the Earth-Sun system, the component of the average torque along the line of equinoxes is
\begin{equation}
\bar{\Gamma}_{x,ES} = \left[ \frac{3GM_{sun}}{2d^{3}}(I_{z'z'}-I_{x'x'})\cos\theta \right].
\end{equation}
\noindent
For the non-magnetic CV DN system, $M=M_{2}$ and is not in solar masses.  For the misaligned CV DN system, the component of average torque along the line of nodes is
\begin{equation}
\bar{\Gamma}_{x,DN} = \left[ \frac{3GM_{2}}{2d^{3}}(I_{z'z'}-I_{x'x'})\cos\theta \right].
\end{equation}
\noindent
Note that the direction is negative as indicated by $\bf{z'}$ $\mathrm{x}$ $\bf{z}$.

\subsection{Retrograde Precession}
A rotating body responds to a torque that has been applied over a small period of time by gaining a small amount of angular momentum, \( \bf{\Gamma} = \frac{d}{dt} \bf{L} \).  The direction of the small gain in angular momentum is parallel to the the applied torque and perpendicular to the spin axis $\bf{z'}$ of the body.  The rotating body responds over this same small time interval by moving its spin angular momentum vector to a new position.  By repeating this process many times over a significantly longer period time, the spin axis of the rotating body $\bf{z'}$ traces out a surface of a cone whose axis coincides with the normal to $\bf{z}$.  The direction of the trace is retrograde to the direction of the spin of the body and orbit as shown in Figure 1.  The angular speed of the trace is the magnitude of the precessional angular velocity.  Thus, to find the retrograde precession we need to find the time rate of change of the angular momentum and equate that to the torque found in Equation (14).   

To establish the total angular momentum, we need to find the total angular velocity.  By definition, the total angular velocity of a body rotating about its axis $\bf{\dot{\Omega}}$ is the sum of the spin rate $N$, which is along the $\bf{\hat{z'}}$ axis, and the motion of its spin axis $\bf{z'}$, which is along the $\bf{\hat{x'}}$ axis,
\begin{eqnarray}
\bf{\dot{\Omega}}&=&N \hat{\bf{z'}} + \left( \hat{\bf{z'}} \mathrm{x} \frac{d\hat{\bf{z'}}}{dt} \right).
\end{eqnarray}
\noindent
In this equation, we assume that no net torque is applied and $I_{x'x'}=I_{y'y'}$.  From Euler's equations of motion, we find 
\begin{equation}
N=\frac{I_{z'z'}-I_{x'x'}}{I_{x'x'}}\dot{\epsilon}.
\end{equation}
\noindent
Upon substitution into Equation (1), the total angular momentum becomes
\begin{eqnarray}
\bf{L}&=&I_{z'z'}N \hat{\bf{z'}}+I_{x'x'} \left(\hat{\bf{z'}} \mathrm{x} \frac{d\hat{\bf{z'}}}{dt} \right),
\end{eqnarray}
\noindent
of which the time rate of change is
\begin{eqnarray}
\frac{d}{dt} \bf{L} & =  & I_{z'z'} N \frac{d\hat{\bf{z'}}}{dt} + I_{x'x'} \left( \hat{\bf{z'}} \mathrm{x} \frac{d^{2}\hat{\bf{z'}}}{dt^{2}} \right).  
\end{eqnarray}
\noindent
In obtaining this last equation, we assume a constant spin rate and constant moment of inertia coefficients.  

The time rate of change of the angular momentum is just the torque we found in the previous subsection.  Even though the magnitude of this torque is not constant over one orbit and therefore the motion of the angular momentum vector moving to its new position pulses with this driving torque, we can assume that over long periods of time this net retrograde precessional angular velocity $\dot{\gamma}$ is steady.  Other than what is mentioned here, we do not consider nodding motions any further in this work.

After setting the time rate of change of the angular momentum equal to the torque found in the previous subsection and after simplifying, we obtain the quadratic
\begin{equation}
\dot{\gamma}^{2} - \left[ \frac{I_{z'z'}N}{I_{x'x'}\cos\theta}\right] \dot{\gamma} + \frac{3GM}{2d^{3}}\frac{(I_{z'z'}-I_{x'x'})}{I_{x'x'}} = 0
\end{equation}
\noindent
whose roots are real and can be summed, the smaller root being
\begin{equation}
\dot{\gamma} = - \frac{3GM}{2d^{3}N}\left(\frac{I_{z'z'}-I_{x'x'}}{I_{z'z'}}\right)\cos\theta.
\end{equation}
\noindent
Like for Equations (15) and (16), we can substitute $M=M_{sun}$ in Equation (22) for the Sun that is located at a distance $d$ from the origin and $M=M_{moon}$ for the Moon that is located at a distance $a$ from the origin 
\begin{equation}
\dot{\gamma} = - \frac{3G}{2N} \left( \frac{M_{sun}}{d^{3}} + \frac{M_{moon}}{a^{3}}\right) \left(\frac{I_{z'z'}-I_{x'x'}}{I_{z'z'}}\right)\cos\theta.
\end{equation}
\noindent
This equation is the retrograde precession of the zero-point of Right Ascension (a.k.a., the First Point of Aries) in the Earth-Moon-Sun system (see any fundamentals of geophysics or celestial mechanics book such as Danby 1962 and Stacey 1977).  

Likewise, we could substitute $M=M_{2}$ for a secondary that is located at a distance $d$ from the origin
\begin{equation}
\dot{\gamma} = - \frac{3GM_{2}}{2d^{3}N}\left(\frac{I_{z'z'}-I_{x'x'}}{I_{z'z'}}\right)\cos\theta.
\end{equation}
\noindent
This equation is the retrograde precession of a point on the rim of the spinning, mis-aligned disk.  This point can be considered equivalent to the zero-point of Right Ascension (a.k.a., the First Point of Aries) in the Earth-Moon-Sun system.  

Because the disk is spinning and because it is tilted, forces that arise from the gradient of the potential are not colinear, and these non-colinear forces cause a net torque on the disk, one component of which is along the line of nodes.  This component of net torque causes the spin axis of the disk to precess in a direction that is retrograde to the orbital direction of the secondary.  The net motion traces out a cone whose axis coincides with the normal to the orbital plane.  We note that this retrograde precessional equation for a spinning tilted accretion disk in a non-magnetic system is generic in that we have yet to take into account the moment of inertia coefficients, a differentially rotating disk, and any effects due to the gas stream on the accretion disk.  

\section{Retrograde Precession in Tilted Disks - Solutions To Theoretical Expressions}
Retrograde precession expressions that have been found by others for various systems can also be found by the technique developed in \S 2, and we show this in this section.  Insight is gained by comparing derivation results.  In all examples, we assume Equation (24) is valid.  In addition, we assume the disk is of constant density, orbits are circular, and Equation (3) is valid for finding the moment of inertia.

\subsection{Rigid, Continuous, Circular, Tilted Disk}
We solve Equation (3) in cylindrical coordinates to find the moment of inertia of a rigid, continuous disk that has constant density.  We also assume the radius of the disk is much larger than the height of the disk, $r_{d} \gg h_{c}$.  For a thin circular disk, 
\begin{equation}
I=\left[ \begin{array}{ccc}
  \frac{1}{4}M_{d}r_{d}^{2} & 0 & 0 \\
  0 & \frac{1}{4}M_{d}r_{d}^{2} & 0 \\
  0 & 0 & \frac{1}{2}M_{d}r_{d}^{2}
         \end{array} \right].
\end{equation}
\noindent 
We assign $I_{x'x'}$, $I_{y'y'}$, and $I_{z'z'}$ to the diagonal elements in the matrix, noting that $I_{x'x'}$=$I_{y'y'}$ and thus $N=\dot{\epsilon}$.  Substituting these into Equations (24), the retrograde precession becomes 
\begin{equation}
\dot{\gamma} = - \frac{3}{4} \frac{G}{\dot{\epsilon}} \frac{M_{2}}{d^{3}} \cos\theta.
\end{equation}
\noindent
Equation (26) is the same result as that obtained by Katz et al. (1982) who approximated a disk as a thin ring to find the nodding motion of accreting rings and disks in X-ray binaries.  In their work, they consider the ring/disk to be tilted and acted upon by a secondary mass.   The Keplerian orbital angular velocity,
\begin{equation}
\dot{\omega}=\sqrt{\frac{GM_{2}}{d^{3}}},
\end{equation}
\noindent
is easily identified in Equation (26).   We assume the angular velocity of the disk around the $\bf{z'}$ axis is $\dot{\omega_{d}}=\dot{\epsilon}$.  Thus, the retrograde precession of a thin, circular, continuous, constant density, rigid, tilted disk is
\begin{equation}
\dot{\gamma} = - \frac{3}{4} \frac{\dot{\omega}^{2}}{\dot{\omega}_{d}} \cos \theta.
\end{equation}
\noindent
At the edge of the continuous rigid disk, the centripetal force equals the gravitational force.  Therefore, we can also represent the angular velocity of the disk at the disk edge by 
\begin{equation}
\dot{\omega}_{d}=\sqrt{\frac{GM_{1}}{r_{d}^{3}}}.
\end{equation}
\noindent
We can substitute this last equation and Equation (27) into Equation (28) to obtain 
\begin{equation}
\dot{\gamma} = - \frac{3}{4} q \left( \frac{r_{d}}{d}\right)^{3} \dot{\omega}_{d} \cos\theta 
\end{equation}
\noindent
where mass ratio $q=M_{2}/M_{1}$.   

If we assume $n=2$ in the polytropic equation of state used by Larwood (1997), then their expression for retrograde precession for protoplanetary accretion disks reduces to our Equation (30).  Likewise, if we assume $n=1.5$ in the polytropic equation of state used by Romero et al. (2000), then the expression they assume for retrograde precession in quasars and black hole accretion disks reduces to our equation.  This equation is proportional to the result obtained by Kondo, Wolff, \& van Flandern (1983), and subsequently by Barrett, O'Donoghue, \& Warner (1988), who establish the precession rate of an orbiting particle in a tilted disk.  Their proportionality constant is 1/2, however.

Substituting Newton's version of Kepler's Third Law for the binary orbit into Equation (30), we obtain yet another version of retrograde precession
\begin{equation}
\dot{\gamma} = - \frac{3}{4} \frac{q}{\sqrt{1+q}} \cos\theta \left(\frac{r_{d}}{d}\right)^{3/2}  \dot{\omega} . 
\end{equation}
\noindent  
If we assume a small angle approximation, then Equation (31) reduces to a later derivation by Warner (1995, 2003) that is based on a retrograde precession of a tilted disk's outer annulus (Kumar 1986, 1989).  The absolute magnitude of Equation (31) is the same result as that obtained by Osaki (1985) for prograde precession in non-tilted or non-warped disks, a result that is too high for prograde precession as pressure effects within the disk have been neglected (Lubow 1992).  

The expressions for retrograde precession developed by others, shown above, are the equivalent to our derivations if we assume that the disk rigidly rotates at the same rate as the primary, i.e., $\dot{\omega}_{d}=\dot{\epsilon}$.  In other words, differential rotation has been neglected.  

\subsection{Differentially Rotating, Circular, Tilted Disk}
In obtaining $I$ in Equation (25), we assume the disk to be circular, thin, of constant density, rigid, and continuous.  However, Papaloizou \& Terquem (1995) note that accretion disks are not solid objects, and thus angular frequency is not constant with radius in fluid disks.  That is, disks are fluid and thus individual rings can rotate differentially and this should be considered in evaluating Equation (25).  

From Equation (25), we note that $I_{z'z'}-I_{x'x'}=I_{x'x'}$.  We assume this same equation holds in a differentially rotating disk.  Equation (24) simplifies to 
\begin{equation}
\dot{\gamma} = - \frac{3}{2} \frac{G}{\dot{\epsilon}} \frac{I_{x'x'}}{I_{z'z'}} \frac{M_{2}}{d^{3}} \cos \theta.
\end{equation}
\noindent 
Angular frequency is not expected to be constant with disk radius and thus Equation (29) is revised to its most general form,
\begin{equation}
\dot{\omega}_{r} = \sqrt \frac{GM_{1}}{r^{3}},
\end{equation}
\noindent
where $r$ is the radial distance from the origin to any particle within the disk. As we still assume the disk is circular, we still expect $I_{x'x'}$=$I_{y'y'}$ even though the disk is differentially rotating.  We do expect that the differential angular velocity to affect the $I_{z'z'}$ coefficient.  With these assumptions and allowing $\dot{\epsilon}=\dot{\omega_{r}}$, the retrograde precession of a thin, circular, constant density, differentially rotating, tilted disk is
\begin{equation}
\dot{\gamma} = - \frac{3}{2} \frac{GM_{2}}{d^{3}} \frac{I_{x'x'} \cos\theta}{G^{1/2}M_{1}^{1/2}J_{z'z'}}
\end{equation}
\noindent
where 
\begin{equation}
J_{z'z'}=\int_{-h_{c}/2}^{h_{c}/2} \int_{0}^{2\pi} \int_{0}^{r_{d}} \rho r^{-3/2} r^{3} dr d\phi dz'.
\end{equation}
\noindent
Note that because the units of $I_{z'z'}$ change when we include the $r^{3/2}$ term inside the integral, the integral is redefined to $J_{z'z'}$.  The solution to this integral is, assuming a constant density disk, $J_{z'z'}=(4/5)\pi h_{c}r_{d}^{5/2}\rho$.  Similar calculations of $J_{x'x'}$ and $J_{y'y'}$ show that our assumption of \( I_{x'x'} = I_{y'y'} \) is correct in Equation (32).  As $\rho=M_{d}\pi^{-1}r_{d}^{-2}h_{c}^{-1}$ for a cylinder, then \( J_{z'z'}=(4/5)M_{d}r_{d}^{1/2} \).  Substituting this result into Equation (34) and substituting in our prevously found solution for $I_{x'x'}=1/4M_{d}r_{d}^{2}$, the retrograde precession of a thin, circular, constant density, differentially rotating, tilted accretion disk is
\begin{eqnarray}
\dot{\gamma} & = & - \frac{15}{32} \frac{M_{2}}{M_{1}} \left( \frac{r_{d}}{d} \right)^{3/2} \left( \frac{GM_{1}}{d^{3}} \right)^{1/2} \cos \theta \\ 
               & = & - \frac{15}{32} \frac{M_{2}}{M_{1}} \left( \frac{r_{d}}{d} \right)^{3} \dot{\omega}_{d} \cos\theta \\
               & = & - \frac{15}{32} \frac{q}{\sqrt{(1+q)}} \cos\theta \left( \frac{r_{d}}{d} \right)^{3/2} \dot{\omega}.
\end{eqnarray}
\noindent
To obtain Equation (37), we have substituted the angular velocity at the disk edge, Equation (29), into Equation (36).   To obtain Equation (38), we have substituted in Newton's version of Kepler's third law.  As we noted earlier, this derivation assumes constant surface density and constant disk radius $r_{d}$.  

Equation (37) is the retrograde precession found by Larwood et al. (1996) for pre-main sequence binary systems where they assume a polytropic gamma of 5/3 (i.e., isentropic) in their equation of state.   If viscous dissipation occurs, then an efficient cooling mechanism is assumed if fluids are to remain isentropic throughout.  Equation (37) is the same result as that found by Terquem, Papaloizou, \& Nelson (1999) for protostellar disks that have tidally induced warps.  Equation (37) is proportional to that found by Larwood (1998) for forced precession in X-ray binaries.  Larwood (1998) found a 3/7 proportionality to our 15/32 as they averaged the tidal torque over the disk.  Equation (38) is from Montgomery (2004).

We note that Equation (35) is not entirely correct.  In this equation, we integrate from the origin to the outer radius of the disk.  The inner rim of the disk does not extend to the origin.  However, the outer radius of the disk is so large compared to the radial extent of the white dwarf or to the radial extent to the inner rim of the disk that the effect on precession is minimal.  Hence, Equation (35) is valid for our study of precession in CV DN systems.

In the derivation of Equations (36)-(38), we assume that the disk differentially rotates.   Implied in this derivation is that the disk precesses as a unit.  Support for this assumption is given by Papaloizou \& Terquem (1995) who analytically show, and Larwood et al. (1996) numerically verify, that rigid body precession of a Keplerian disk is possible so long as the sound crossing time scale in the disk is small compared to the precession time scale of the disk.  Further, numerical simulations by Murray (1998) and observations by Patterson et al. (1998) do not indicate that disks differentially precess.  

\subsection{Differentially Rotating, Elliptical, Tilted Disk}
In \S 2.4, we assume that the major components in accreting CV DN systems are the primary, secondary, and the accretion disk.  We assume that the orbit is circular, symmetry is about the $\bf{z'}$ rotation axis, and the disk is circular.  A circular disk is mirror symmetric and rotationally symmetric around its $z'$ axis and, as a result, we could find equal body moment of inertia coefficients $I_{x'x'} = I_{y'y'}$ and a constant rotation around the normal axis to the plane of the disk, Equation (24).  

However, accretion disks in CV DN systems have long been thought to be non-circular (see e.g., Warner 1995, 2003).  An elliptical disk does not have equal radii.  Thus, the moment of inertia coefficients about the principle axes are not equal, $I_{x'x'}$ $\ne$ $I_{y'y'}$, and $\Gamma_{z} \ne 0$.  Yet, over a time period of one orbit, $\bar{\Gamma}_{x}$ still dominates.  Equations (31) and (38) are still valid in non-circular disks so long as the disk precesses as a unit, the distance $d$ is much greater than the radius of the primary as discussed in \S 2.2, and precession time is much longer than the orbital time.

\subsection{Rotating, Circular, Tilted Ring}
In the derivation of Equations (36)-(38), we assume that the disk differentially rotates yet the disk precesses as a unit.  We have justified the geometry of the disk in neglecting a small, centrally located inner hole where the primary is located and in neglecting the outer shape of the disk being more elliptical than circular in CV DN non-magnetic systems.  We have taken into account major components of the system - the primary, secondary, and the accretion disk.  However, we have neglected the effect of the gas stream on the accretion disk.  Therefore, we introduce additional geometrical shapes that may be used to help describe the overall geometrical shape of the disk as the overall geometrical disk shape may have changed due to the gas stream striking the disk.  

In this subsection, we consider the geometrical shape of a ring that is centered around the primary and that is tilted at the same angle as the disk.  The inner rim radius of the ring $r_{inner}$ is located near the innermost rim of the disk and the outer rim radius of the ring is located at $r_{outer}$. In this derivation, we also assume rotation and precession as a unit.  

In modeling black holes, Bardeen \& Petterson (1975) state that the combined effects due to precession and viscosity should be to align the rotation of the inner disk annuli with the spin axis of the hole.  Kumar and Pringle (1985) show that the alignment of the disk and the hole does occur in slightly tilted, steady state disks.  Borrowing these ideas, we return to Equation (32) and assume  that $\dot{\epsilon}=\dot{\omega}_{inner}$
\begin{equation}
\dot{\gamma} = - \frac{3}{2} \frac{G}{\dot{\omega}_{inner}} \frac{I_{x'x'}}{I_{z'z'}} \frac{M_{2}}{d^{3}} \cos \theta.
\end{equation}
\noindent
where 
\begin{equation}
\dot{\omega}_{inner}=\sqrt{\frac{GM_{1}}{r_{inner}^{3}}}
\end{equation}

\noindent
is the angular velocity at $r_{inner}$.  Note that we assume that the inner edge of the ring rotates at the same rate as the primary.  Using Kepler's Third Law, we can relate the radius and angular velocity of the inner ring edge to that of the outer ring edge
\begin{equation}
r_{inner}=r_{outer}\left( \frac{\dot{\omega}_{outer}}{\dot{\omega}_{inner}} \right)^{2/3}
\end{equation}
\noindent
to obtain
\begin{equation}
\dot{\gamma} = - \frac{3q}{2} \left( \frac{r_{inner}}{d}\right)^{3/2} \frac{I_{x'x'}}{I_{z'z'}} \frac{(GM_{1})^{1/2}}{d^{3/2}} \cos \theta.
\end{equation}
\noindent
Substituting in Kepler's Third Law for the orbital period of the binary orbit, Equation (41), and the moment of inertia coefficients from Equation (25), Equation (42) becomes
\begin{equation}
\dot{\gamma}  = - \frac{3}{4} \frac{q}{\sqrt{1+q}}  \cos \theta \left( \frac{r_{outer}}{d}\right)^{3/2} \dot{\omega} \left(\frac{\dot{\omega}_{outer}}{\dot{\omega}_{inner}}\right) 
\end{equation}
\noindent
or equivalently
\begin{equation}
\dot{\gamma}  = - \frac{3}{4} q \left( \frac{r_{inner}}{d}\right)^{3}  \dot{\omega}_{inner} \cos \theta  
\end{equation}
\noindent
or equivalently
\begin{equation}
\dot{\gamma} = - \frac{3}{4} \frac{q}{\sqrt{1+q}}  \cos \theta \left( \frac{r_{inner}}{d}\right)^{3/2} \dot{\omega}.
\end{equation}
\noindent
Notice the similarities between Equations (30) and (44) and between Equations (31) and (45).  

To reduce Equation (45) further, we need to know the radius to the inner rim of the circular ring.  The inner rim radius of the ring is unlikely to be the radius of the white dwarf as the inner rim of the ring is unlikely to extend down to the surface of the primary.  Also, the inner rim radius of the ring is not likely to be the distance of closest approach of a free-falling gas stream to the primary (Lubow \& Shu, 1975)
\begin{equation}
r_{min}=0.0488dq^{-0.464},
\end{equation}
\noindent
which is good for \( (0.03 < q < 1) \).  Although this distance of closest approach yields a radius that is larger than the radius of the white dwarf, this radius is not large enough to explain a circularized ring:  The initial path of the free-falling gas stream forms a parabolic shape around the primary, and after interaction with the primary's Roche lobe, the path then forms a rosette pattern.  Upon collisions of the gas stream with itself, the rosette pattern of the gas stream settles into a circularized ring that has a radius that is larger than the distance of closest approach of the initial free-falling gas stream.  Yet the inner rim radius of the ring cannot be the circularization radius (Hessman \& Hopp, 1990)
\begin{equation}
r_{circ}=0.0859dq^{-0.426},
\end{equation}
\noindent
which is good for \( (0.05 \le q < 1) \), as this equation describes the $\emph{minimum}$ radius of the ring's $\emph{outer}$ radius.  

As the ring is thin, we shall assume that $r_{inner}$ is half way between the minimum radius and the circularization radius
\begin{eqnarray}
 r_{inner} & = & \left( \frac{r_{min}+r_{circ}}{2} \right).  
\end{eqnarray}
\noindent
As Warner (2003) finds that \( r_{circ} \sim 1.75r_{min} \), $r_{inner}  \sim 1.38 r_{min}$ and Equation (45) becomes 
\begin{eqnarray}
\dot{\gamma} & = &- \frac{3}{4} \frac{q}{\sqrt{1+q}}  \cos \theta \left( \frac{1.38r_{min}}{d}\right)^{3/2} \dot {\omega} \nonumber \\
             & = &  - \frac{3}{4}  \frac{q}{\sqrt{1+q}} \cos \theta \left(0.0673q^{-0.464}\right)^{3/2}  \dot{\omega}.
\end{eqnarray}
\noindent
In summary, we assume a thin ring that is centered about the origin and is located near the inner edge of the disk.  This ring retrogradely precesses as a unit.  We note that we could have repeated the process of differential rotation that we introduced in \S 3.2.  However, because the ring is thin, we find that Equation (49) is minimally affected.  

\subsection{Negative Superhump Period Excesses}
As the orbital period is slightly longer than the period of the secondary encountering the same disk node per orbit, we can relate the negative superhump period $P_{-}$ and the orbital period to the retrograde precession period by 
\begin{equation}
P_{r}^{-1} = P_{-}^{-1} - P_{orb}^{-1}.  
\end{equation}
\noindent
Another relation is the negative superhump period excess 
\begin{eqnarray}
\epsilon_{-} & = & 1 - \frac{P_{-}}{P_{orb}}    \\
	     & = & 1 - \left[ 1 + \frac{P_{orb}}{P_{r}} \right]^{-1} 
\end{eqnarray}
\noindent
where we substitute Equation (50) into Equation (51) to obtain Equation (52).  For example, we can find the negative superhump period excess for the rigid, continuous, circular, tilted disk from \S 3.1.  If we assume that the radius of the disk is that given by Paczynski (1977) 
\begin{equation}
r_{d}=\frac{0.6d}{1+q},
\end{equation}
\noindent
which is valid for mass ratios in the range \( 0.03<q<1 \), then we can combine Equations (31) and 
(52) to obtain
\begin{eqnarray}
\epsilon_{-,rigid} & = & 1 - \left[ 1 + \frac{0.349q \cos \theta}{(1+q)^{2}} \right]^{-1}.
\end{eqnarray}

Likewise, we can find the negative superhump period excess for the differentially rotating, circular, tilted disk from \S 3.2 by substituting Equation (53) into Equation (38) and then into Equation (52) to obtain
\begin{eqnarray}
\epsilon_{-,disk} & = & 1 - \left[ 1 + \frac{0.218q \cos \theta}{(1+q)^{2}} \right]^{-1}.
\end{eqnarray}

Likewise, we can find the negative superhump period excess for the rotating, circular, tilted ring from \S 3.3 by substituting Equation (49) into Equation (53) and then into Equation (52) to obtain	    
\begin{equation}
\epsilon_{-,ring}   = 1 - \left[ 1 + \frac{\frac{3}{4}q \cos \theta}{\sqrt{(1+q)}} \left(0.0673q^{-0.464}\right)^{3/2}\right]^{-1}.
\end{equation}
\noindent
Note that we have dropped the negative sign that redundantly indicates retrograde precession.  

Equation (55) appears in Montgomery (2004).  If we assume a primary mass $M_{1}$=0.8$M_{\odot}$ and mass ratio $q=0.4$, then the secondary mass $M_{2}$=0.32$M_{\odot}$.  The radius of the disk is $r_{d}\sim0.43d$, where $d\sim1.23R_{\odot}$ (Montgomery 2009).  The rotation period at the disk edge, determined from Equation (29), is around one hundred seconds.  This rate is fast compared with the orbital period $P_{orb}=3.57$ hours found using Newton's version of Kepler's third law.  From Equation (55), we find $P_{r}\sim22.5P_{orb}=$80 hours, assuming small angle approximation for disk tilt.  

Instead of mass ratio, we can express Equations (54), (55), and (56) in terms of the orbital period.  
For this substitution, we shall assume non-magnetic, accreting binaries that follow the Smith \& Dhillon (1998) secondary mass-period relation, 
\begin{equation}
M_{2}/M_{\odot}=(0.038 \pm 0.003)P_{orb}^{1.58\pm0.09} 
\end{equation}
\noindent
where $P_{orb}$ is in hours.  We do not show these lengthy negative superhump period excess equations in this work, but they can easily be found by variable substitution where $q=\frac{M_{2}}{M_{1}}$ and $M_{1}=0.8M_{\odot}$.  

\section{Numerically Simulated Retrograde Precessions}
In Montgomery (2009), we simulate non-magnetic CV DN systems that have mass ratios within the range \( (0.35 \le q \le 0.55) \) and low mass transfer rates.  By limiting our mass ratio range to longer orbital periods, we eliminate systems that apsidally precess and thus show positive superhumps in their light curves, thereby minimizing variables.   We adopt a Smith \& Dhillon (1998) secondary mass-period relation which is for main-sequence dwarfs as the secondaries.   Because we simulate accreting, non-magnetic CV DN systems in or above the period gap, we choose an average white dwarf primary mass $M_{1}=0.8M_{\odot}$, the average value found by Smith \& Dhillon (1998) for this mass ratio range.  We refer the reader to Montgomery (2009) for details regarding the code yet we provide a summary here.

For our numerical simulations, we use Smoothed Particle Hydrodynamics (SPH) to simulate accretion disks that have constant and uniform mass particles.  In the code, we assume assume a constant and uniform smoothing length $h$ and an ideal gamma-law equation of state \( P= (\gamma-1) \rho u \) where $P$ is pressure, $\gamma$ is the adiabatic index, $\rho$ is density, and $u$ is specific internal energy.  The sound speed is  \( c_{s} = \sqrt{\gamma (\gamma -1) u}\).  The code utilizes a Lattanzio et al. (1986) artificial viscosity, and we choose \( \alpha = \beta = 0.5 \) viscosity coefficients.  Our viscosity is approximately equivalent to a Shakura \& Sunyaev (1973) viscosity parametrization (\(\nu = \alpha' c_{s} H \) where $H$ is the disk scaleheight) $\alpha$'=0.05.  As Smak (1999) estimates $\alpha' \sim $ 0.1 - 0.2 for DN systems in high viscosity states, our simulations are more for quiescent systems.  As only approaching particles feel the viscous force and since neither radiative transfer nor magnetic fields are included in this code, all energy dissipated by the artificial viscosity is transferred into changing the internal energies.  

As radiative cooling is not included in this code, an adiabatic index $\gamma$=1.01 is incorporated to prevent internal energies from becoming too large.  By integrating the changes in the internal energies of all the particles over a specific time interval $n$, variations in the bolometric luminosity yield an approximate and artificially generated light curve 
\begin{equation}
L_{n} = \sum_{i} \mathrm{du_{i}^{n}}.
\end{equation}
\noindent
In all simulations, we build a disk to 100,000 particles.  At orbit 200, we artificially rotate the disk out of the orbital plane 5$^{o}$ to induce negative superhumps in the light curve and to force the disk to precess in the retrograde direction.  For example, if $q=0.4$ then $M_{2}$=0.32 $M_{\odot}$.  Our simulation unit length scales to \( d \sim 1.23 R_{\odot} \) if we assume the secondary mass-radius relationship \( R_{2}=(M_{2}/M_{\odot})^{-13/15} R_{\odot} \) (Warner 2003) that applies for \( 0.08 \le M_{2}M_{\odot}^{-1} \le 1.0 \) and the Eggleton (1983) volume radius of the Roche lobe secondary
\begin{equation}
\frac{R_{2}}{a} = \frac{0.49 q^{2/3}}{0.6 q^{2/3} + \ln(1 + q^{1/3})}.
\end{equation}
\noindent
In this equation, $R_{2}$ and $M_{2}$ are in solar radii and solar mass, respectively.  The Eggleton (1983) relation is good for all mass ratios, accurate to better than 1\%.  Taking the radius of the primary to be 6.9x10$^{8}$ cm, the scaled radius of the white dwarf is $R_{1}=0.0081d$.  

After building the disk, we allow the disk to evolve in the short term to a quasi-equilibrium state where particles are injected at the rate they are removed from the system by either being accreted onto the primary or the secondary or lost from the system.  The average net rate of accretion for a quasi-static disk is around 500 particles per orbit as shown in Montgomery (2009).  As this net rate is approximately half that injected to build the disk, the steady state mass transfer rate reduces to $ \dot{m} \sim$ 1 x $10^{-10} M_{\odot}$/yr or a rate similar to an SU UMa in quiescence.  We estimate the density of the gas near $L_{1}$ to be $\rho \sim 10^{-10}$ g cm$^{-3}$ using \( m_{p}=(4/3)\pi h^{3} \rho \).   As we maintain 100,000 particles in the steady state disk, then the total mass of the disk is  $ M_{d} $= 100,000 x $m_{p}$ = 3.5x10$^{22}$ g or \( M_{d} \sim \) 2x10$^{-11} M_{\odot}$, a negligible value compared to the mass of either star.  If we assume the outer radius of the disk is \( R_{d} \approx 2 r_{circ} \), or twice the circularization radius (Warner 2003)
\begin{equation}
\frac{r_{circ}}{a} = 0.0859 q^{-0.426}
\end{equation}
\noindent
where \( 0.05 \le q < 1 \), then for $q=0.4$ we find \( r_{circ} \sim 0.13 d \) and \( R_{d} \sim 0.25 d \).   As \( \rho << 
M_{d} R_{d}^{-3} \), self gravity is neglected in these simulations.

In Montgomery (2009), we find the calculated negative superhump period, its propagated error, and the nodal superhump period excess as output to the code.  We list these values in Table 1.  The units for period are hours. We also find the source that powers the negative superhump.  Our results suggest that negative superhumps are due to an increase in the number of particles that reach innermost annuli of a tilted disk (Montgomery, 2009).   As inner annuli have higher energies, the combination of increased number of particles and higher energy inner annuli yields more light being emitted.  This additional light waxes and wanes with the amount of gas stream overflow and the amount of particle migration into innermost annuli.  As the secondary orbits the tilted disk, the gas stream mostly oveflows one face of the tilted disk for approximately the first half of an orbit and then over the other face of the tilted disk for approximately the second half of an orbit.  As a result of the disk tilt and thus gas stream overflow and additional particle migration to innermost annuli, a yellow colored ring that is denser that its surroundings is generated near the innermost annuli (see Figure 3).  This denser ring is important to the generation of the negative superhump.   

We do note that the relative surface densities across the face of the disk vary by only $\sim4\%$ from annuli to annuli, regardless if the disk is tilted or not.  Exceptions are the disk rim and a denser ring, colored green, located near $\sim0.2$d in Figure 3.  The latter is caused by gas stream overflowing the tilted disk and striking the disk face.  Of the two accretion disks shown in Figure 3, the tilted disk is more blended in density.  Hence, assuming a constant density disk in our derivations of retrograde precession in $\S3$ is justified. 

Note in Figure 3 that the ring is tilted at the same angle as the disk and that the ring is rotating.  Therefore, in addition to the disk, the ring is also subject to precession.  Also note that numerically simulated CV DN disks (Figure 3) are elliptical and do not have excessively large, centrally located inner holes.  As the disk radius is much larger than the radius of this small hole, our previous assumption of integration from the origin of the system to the outer edge of the disk is justified.  

\section[]{OBSERVATIONAL DATA}
Table 2 lists observational data including orbital periods, negative superhump periods, and negative superhump period excesses for several non-magnetic, accreting CV DN systems.  If errors are known, they are listed in parentheses.  In the table, NL is novalike, PS is the permanent superhumper category of NL, IP is intermediate polar, SW Sex and VY Scl are members of NLs, ER UMa is a member of the SU UMa subclass. 

In Table 2, we also list positive superhump periods and precessional data for those systems that precess both in the prograde and in the retrograde direction.  Prograde precession is due to density waves forming radially through the middle-to-outer annuli of the disk (Smith et al. 2007) and is described by a dynamical prograde term plus a retrograde pressure term (Lubow 1991a,b).  The retrograde precession term is a correction to the prograde precession term due to these spiral density waves.  In this work, we do not establish the geometrical shape of an accretion disk with these radial density waves as these radial density waves are not generated in long orbital period CV DN systems.  We leave this as future work.  

\section[]{COMPARISONS WITH OBSERVATIONS}
Figure 4 shows a comparison of the observational nodal superhump period excess ($\square$ for IPs, $\diamond$ for AM CVn, * for CV DN that progradely and retrogradely precess, and X for those systems that only retrogradely precess) to the numerical negative superhump period excess ($\triangle$) values listed in Tables 1 and 2, respectively, as a function of orbital period.  The width of the symbol can be taken as the error in the orbital period.  In addition, we overplot a short dashed line that represents a retrogradely precessing rigid disk - Equation (54), a dash-dot line that represents a retrogradely precessing but differentially rotating disk - Equation (55), a long dash line that represents a retrogradely precessing ring - Equation (56), and a solid line that represents a disk that is differentially rotating yet retrogradely precessing as a unit and has a precessing ring.  

As shown in Figure 4, the retrograde precession of an inner annuli ring (long dashed line) does not match any of the observations, a result in agreement with the idea that the disk cannot be represented by a ring.  The retrograde precession of solid, rigid disk also does not match any of the observations, a result in agreement with the idea that the disk rotates differentially as the disk is a fluid.  

Also shown in Figure 4, the retrograde precession of a differentially rotating disk does not match the numerical simulations.   Of the four lines shown in Figure 4, the numerical simulations are best described by the solid line.  To generate the solid line in Figure 4, we assume that the net retrograde precession is due to a differentially rotating disk that has superimposed on it a rotating ring due to the gas stream overflowing the tilted disk and particle migration as the secondary orbits.  The net precession is
\begin{equation}
\frac{P_{orb}}{P_{r,net}} =  Equation (38) + Equation (49) .
\end{equation}
\noindent
Likewise, the net negative superhump period excess is
\begin{equation}
\epsilon_{-,net} =  Equation (55) + Equation (56)
\end{equation}
\noindent
as shown in Figure 4. 

Smith \& Dhillon (1998) find that CV secondary stars with orbital periods shorter than 7-8 hours are indistinguishable from main-squence stars in detached binaries in terms of spectral type, mass, and radius.  They find that CVs with periods up to 7-8 hours almost always have main-sequence secondaries, although they do point out that systems with peculiar secondaries do exist.  We assume their secondary mass-period relation in both our theoretical expressions and in our numerical simulations.  If Smith \& Dhillon (1998) and this work are correct, then the solid line in Figure 4 should represent those systems that have main sequence secondaries.  The one observational system that is on the solid line is V1193 Ori.  HS 1813+6122 is nearby to one of our numerical simulation points.  As Smith \& Dhillon (1998) find that the average white dwarf mass above the period gap is $M_{1}=0.80\pm0.22 M_{\odot}$, the same average mass we chose for the numerical simulations and  theoretical expression for retrograde precession, we find $M_{2}\sim0.267M_{\odot}$ and $q\sim0.33$ for V1153 Ori.  We find $M_{2}\sim0.281M_{\odot}$ and $q\sim0.35$ for HS 1813+6122.  In agreement with the ideas of Smith \& Dhillon (1998), these above-the-period-gap systems have stellar masses.  

We do note the large scatter in the observational data.  As shown in Figure 4 by the various lines, the geometric shape of the disk significantly affects the precession rate and could be a major source to the scatter.  Another smaller source to the scatter not considered in this work is density waves that may be generated in the disk.  Therefore, we have identified the observational data in Figure 4 by various different symbols to indicate systems with different geometries or otherwise unusual disks.  For example, those systems in Table 2 labeled IPs are expected to have a large absence of inner annuli due to the primary's large magnetic field that disrupts the disk.  In this work, we did not find the retrograde precession of a differentially rotating, retrogradely precessing, hollow cylinder and thus we flagged these observational systems in Figure 4 as squares.  A hole as large as that experienced by an IP's disk would not be described by the solid line or any other line in Figure 4 as the geometrical shape of this system is different than that described by these lines.  Likewise, the geometrical shape of an accretion disk that both progradely and retrogradely precesses would not be described by any of the lines in Figure 4.  These systems have radial density waves in their accretion disks and may also have an innermost ring due to gas stream overflow and particle migration to innermost annuli because the disk is tilted.  We did not consider these geometrical shapes or disks with density waves in this work.  We leave these different geometrical shape disks for future work. 
 
\section{DISCUSSION}
\subsection{Warp or Tilt?}
Warps in disks can be localized, say in inner regions or in outer regions (e.g., Boffin et al. 2003), or across the entire disk.  An example of localized warps are those generated in inner disk annuli.  Nelson \& Papaloizou (1999) find inner disk warps that are induced by misaligned angular momentum vectors of the disk and the rotating black hole.  Petterson (1977) and Iping and Petterson (1990) study radiation pressure as a source to disk warp.  If a primary is a source of strong radiation, then the surface of the accretion disk can absorb this radiation and re-emit it normal to the disk surface, inducing a torque on the disk that may warp the disk if the central source luminosity exceeds a critical value (see e.g., Petterson 1977; Pringle 1996, 1997; Wijers \& Pringle 1999).  The strong radiation source twists the disk so that each annuli  affected precesses at the same rate.  In these studies, the warp precesses relative to the inertial frame as a rigid body.   Bisikalo et al. (2004) study CV accretion disks and they too find a localized inner disk warp that precesses yet the disk as a whole does not precess.  

An example of entire disk warp that precesses as a unit are the studies of Larwood et al. (1996), Larwood \& Papaloizou  (1997), and Larwood (1997).  These authors find precession rates due to a tidal interaction between a circumprimary disk or a circumbinary disk and a secondary point mass on a parabolic trajectory in protostellar systems.  Their results indicate that disks become modestly warped ($1^{o}-6^{o}$) during non-coplaner encounters and that disks precess approximately as a unit.  Murray et al. (2002) find that a warp can be induced in CV accretion disks by an inclined magnetic dipole field that is centered on the secondary.  They also find that their warp precesses retrogradely and as a unit.  

An example of a localized warp that may cause the entire disk to warp is the studies by Foukes, Haswell, \& Murray (2006) who generate warps in X-ray binary disks.  Their source to warp is also radiation driven by the primary.  They find that inner disk annuli can warp and the warp precesses nearly as a rigid body in the retrograde direction.  For less extreme mass ratio systems, they find that inner and outer regions of the disk warp to the point that they can say that the entire disk is tilted to the binary plane.  As a result of the entire disk tilt, the whole disk precesses in the retrograde direction, a result also found in Montgomery (2004, 2009).  

For CV DN systems, warps seem unlikely and are usually not considered (e.g., Boffin et al. 2003) as  CV DN primary are not known to be strong radiation sources.  In addition, this study considers only CV DN that are considered non-magnetic.  The geometry suggested in Larwood et al. (1996), Larwood \& Papaloizou (1997), and Larwood (1997) is unlikely for CV DN systems.  Also unlikely is a misaligned disk and rotating primary angular momentum vectors as this source seems to only warp inner annuli without the generation of negative superhumps (Foulkes, Haswell, \& Murray 2006).  Disk tilts have been postulated to involve the entire disk or only a portion such as the outer edge (Bonnet-Bidaud, Motch, \&  Mouchet 1985; Kumar 1986, 1989).  As Foulkes, Haswell, \& Murray (2006) and Montgomery (2009) find that whole disk needs to be tilted in order for the gas stream to flow over and under the disk edge, entire disk tilts are the likely geometry for non-magnetic, accreting, tilted, spinning systems if negative superhumps of the correct shape and frequency per orbit are expected to be seen in the light curves.  The source to disk tilt in non-magnetic systems remains unknown, and we leave this as future work.  As Figure 5 shows, a fully tilted disk may look like a disk warp as the system retrogradely precesses.  In Figure 5, we only show one half of the total disk precessional period.

\subsection{Whole or Partial Disk Precession?}
In Montgomery (2009), we find that if the entire disk is tilted, then the gas stream can overflow a disk face and reach inner annuli.  As the secondary orbits, the gas stream flows over one face of the disk for $\sim$1/2 an orbit and the flows over the other face of the disk for $\sim$1/2 an orbit.  The location where the gas stream transitions on the disk rim from flowing over to under and vice versa (i.e., the nodes) precesses in the retrograde direction.  Like the slow Westward movement of the First Point of Aries along the Equator of the Earth, the line of nodes also has a slow Westward movement along the rim of the disk.  If the disk precesses as a unit and the line of nodes shifts Westward by a small amount each orbit, then the location where the gas stream strikes the innermost annuli must also shift Westward by a small amount each orbit.  That is, the spinning, tilted ring formed near the innermost annuli of the disk and formed by the gas stream overflowing the tilted disk also precesses in the retrograde direction.  Therefore, to explain precession in our numerical simulations, we net the retrograde precession of the tilted, rotating ring and the tilted, rotating disk.

Bisikalo et al. (2004) finds that only their spiral density wave retrogradely precesses, not the entire numerical simulated CV accretion disk.  Observations by Patterson et al. (1998) do not indicate that disks precess differentially.  Numerical simulations by Murray (1998) do not indicate that disks precess differentially.  Our results do not indicate that disks precess differentially.  Papaloizou \& Terquem (1995) analytically show, and Larwood et al. (1996) numerically verify, that rigid body precession of a Keplerian disk is possible so long as the sound crossing time scale in the disk is small compared to the precession time scale of the disk.  We find in Montgomery (2009) that the entire disk needs to tilt for additional particle migration to reach inner annuli and produce negative superhumps.  We find in this work that both outer and inner annuli precess in fully tilted disks.  These results suggest that partially retrogradely precessing non-magnetic CV DN tilted accretion disks are unlikely.  

\subsection{Precession Source}
Driven precession involves torques on a body.  One example is a wobbling coin on a table.  The coin spin rate is slow but its wobble rate is fast due to a torque on the body.  Another example is a spinning top inclined at an angle from the vertical.  The spin rate is fast but the precession rate is slow due to torques on the body.  The precession of the Earth due to tidal torques by the Moon and the Sun is another example of a slow precession rate compared to Earth's (relatively) fast spin rate.   For a CV DN system with $q=0.4$, we previously calculated the retrograde precessional period to be approximately 80 hours.  This period is slow compared to the primary's rotation period of around one hundred seconds, if we assume that the primary's rotation period is similar to that at the disk edge.  Therefore, driven precession could be a viable source of retrograde precession in CV DN systems that show negative superhumps in their light curves.  Katz (1973),  Roberts (1974), Katz et al. (1982), Kondo et al. (1983), Kumar (1986, 1989), Barrett et al. (1988), Warner (1995, 2003) are all examples of those employing or assuming torques to induce retrograde precession.  

As we show in Figure 4, retrograde precession in CV DN disks is likely due to the same source that causes the Earth to retrogradely precess, that is, by tidal torques.  The difference in precessional values of the Earths' equinoxes and a CV DN's line of nodes is due to the geometric shape of the precessing object (e.g., oblate spheroid, disk, etc.), the state of matter (i.e., fluid, solid, etc.), and any effects due to the accretion stream.

Bisikalo et al. (2004) generate a spiral density wave in innermost annuli of their numerical simulated CV disks.  Their spiral density wave is caused by sheared elliptical inner disk annuli.  They find that only the spiral density wave retrogradely precesses, and they cite Kumar (1986) as the source to retrograde precession.  However Kumar (1986) finds the retrograde precession due to a tilted disk's outer annuli and thus cannot be the source to retrograde precession of the spiral density wave.  However, we may be able to help Bisikalo et al. (2004).  Lubow \& Shu (1975) find that pressure can be generated in the disk due to shocks that are induced by gas stream overflow, and Lubow (1991a,b) finds that additional pressure due to spiral density waves could cause the disk to retrogradely precess.  If the inner disk has a spiral density wave instead of a dense ring, then the net retrograde precession can still be explained by the solid line in Figure 4.  That is, the disk would retrogradely precess due to combined effects due to tidal torques and pressure from a spiral density wave located near the innermost disk annuli (instead of a ring).  We plan to show this in a future work.

\section{SUMMARY, CONCLUSIONS, AND FUTURE WORK}
Negative superhumps are found in observational systems with high mass ratios, long orbital periods, and high mass transfer rates.  In Montgomery (2009), we find that the negative superhump can be due to extra light being emitted from inner disk annuli.  The extra light is modulated by gas stream overflow that strikes inner disk annuli and by particle migration as the secondary orbits.  For this result to be true, we find in Montgomery (2009) that the disk must be fully tilted.  The source to disk tilt has not yet been resolved and we leave this as future work.  Fully tilted disks could be subject to driven torques that cause the disk to retrogradely precess.  

Patterson et al. (1993) suggest that the retrograde precesison in CV DN accretion disks may be by the same source that causes the Earth to retrogradely precess.  Using the Earth-Moon-Sun system as a model, we generate a generic theoretical expression for a spinning disk that is misaligned with the orbital plane to test this idea.  We assume the potential of MacCullogh is valid for our spining, tilted disk that contains a massive primary that is smaller than the thickness of the disk and the distance to the secondary.  Our expression matches that which describes the precession of Earths' equinoxes.

By making appropriate assumptions, we reduce our generic expression for retrograde precession to those generated by others, or to those used by others, to describe retrograde precession in protostellar, protoplanetary, X-ray binary, non-magnetic accreting CV DN, quasar and black hole systems.  Because our generic expression reduces to those generated by others for a variety of systems, retrograde precession by tidal torques on a spinning, misaligned primary may be ubiquitous.  Previous expressions treated the disk or ring as rigid and/or didn't include the effects on the disk by the accretion stream.  Because the disk is fully tilted, the gas stream overflows one face of the disk and particles migrate to inner annuli, creating a spinning, tilted ring that is denser than its surroundings as shown in our numerical simulations.  

In this work, we generate a theoretical expression for the retrograde precession of a spinning, tilted ring.  We assume that the rotation rate of the primary matches that of the inner rim of the spinning, tilted ring.  This implies that the axes of the ring and the primary align.  We show that the net retrograde precession in CV DN non-magnetic accretion disks needs to include the effects due to both the precessing disk and the precessing ring.  If we include the net moment of inertia due to the disk and the ring, then our theoretical expression agrees well with our numerical results.  Our findings support the suggestion by Patterson et al. (1993) that tidal torques like those by the Moon and the Sun on the spinning, tilted Earth are the source to retrograde precession in spinning, tilted CV DN accretion disks.  

In our theoretical expression and our numerical simulations, we assume an average primary mass $M_{1}=0.80M_{\odot}$ that matches the average primary mass $M_{1}=0.80\pm0.22 M_{\odot}$ found by Smith \& Dhillon (1998).  If the results in Smith \& Dhillon (1998) and this work are correct, then from Figure 4 we find that V1193 Ori and HS 1813+6122 secondaries may have stellar masses.  

The scatter in observational precessional values seems to be mostly from differing geometrical disk shapes.  The scatter could also be from any density waves that may be present in the disk.  Bisikalo et al. (2004) finds that a spiral density wave is generated in innermost annuli of their numerically simulated CV accretion disks.  They find that their spiral density wave retrogradely precesses.  Pressure effects generated within the disk from a spiral density wave could add to retrograde precession.  The disk could retrogradely precess due to combined effects due to tidal torques and pressure from a spiral density wave.  We plan to show this in a future work, and we anticipate that the result is the same solid line shown in Figure 4.  However, unlike that found by Bisikalo et al. (2004), we expect the disk to still precess as a unit. 

We do note that we did not consider significant mass loss of the secondary over time and this may also affect precessional values.  If the mass lost by the secondary is not significantly acquired by the primary and is instead mostly lost via e.g., winds, jets, or  circumbinary disks, then the geometric shape of the disk, moment of inertia, mass ratio, and/or orbital period changes and thus so changes the precessional values.   Whether the retrograde precession and orbital period values change one-to-one is not fully understood and we leave this as future work as well. 

\section*{Acknowledgments}
The author would like to acknowledge an AAS Small Research Grant and a UCF-UF Space Research Initiative that enabled us to generate this work on our eight-node server.  We would like to thank undergraduate Mark Guasch for colorizing the disks for this work and undergraduate Jon Edmiston for finding the elements of the moment of inertia tensor for a rigid elliptical disk.  We thank Graham Wynn and Luigi Stella, the reviewer and Scientific Editor for ApJ, respectively, as their comments enabled a manuscript that flowed better than the original version.  We especially thank Graham Wynn for his suggestion of differentiating potentially different geometrical disk shapes with different symbols and addressing other geometrical shape disks (e.g., IP disks) that are not considered in this work.

\clearpage

%%FIGURE 1 %%%%%%%%%%%%%%%%%%%%%%%%
\begin{figure}
\epsfysize 2.5in
%\center{\epsfbox{Fig1a.eps}}
%\center{\epsfbox{Fig1b.eps}}
\caption{Sketch of the Earth-Sun binary system (top panel) and of the non-magnetic CV DN system (bottom panel).  Thin solid line represents the axes of the ecliptic with rotation in the prograde, counter-clockwise direction around the z-axis with speed $\dot{\omega}$.  The primary (i.e., the Earth in the top panel or the disk it the bottom panel) is at the origin and rotates counter-clockwise with speed $\dot{\epsilon}$ around the dashed z' axis where (x',y',z') are the body axes.  The z' axis is at an obliquity angle $\theta$ relative to the z-axis.  Note that the primary in the CV DN system (bottom panel) is small compared to the size of the disk and is thus hidden from view.  The net torque around the z-axis is caused by differing gravitational attraction forces between opposite points on the surface of the Earth (top panel) or disk (bottom panel) and the Sun (top panel) or secondary mass (bottom panel), shown as a sphere on the y-axis at a distance $d$ from the origin.  The net torque causes the rotation axis of the primary, z', to precess in the retrograde direction. Schematics are not drawn to scale.}
\label{Figure 1.}
\end{figure}
%%%%%%%%%%%%%%%%%%%%%%%%%%%%%%%%%%%

\clearpage

%%FIGURE 2 %%%%%%%%%%%%%%%%%%%%%%%%
\begin{figure}
\epsfysize 2.5in
%\center{\epsfbox{Fig2.eps}}
\caption{Schematic of a non-spherical attracting body $m$ that has a small element $dm$ located at distance $r'$ from its center of mass and at an angle $\alpha$ from the line of centers connecting the center of mass of the attracting body with the center of mass of a far away point mass body $M$.  The far-away body is located at a distance $d$ where $d>>r'$.  Schematic is not to scale.}
\label{Figure 2.}
\end{figure}
%%%%%%%%%%%%%%%%%%%%%%%%%%%%%%%%%%%

\clearpage

%%FIGURE 3 %%%%%%%%%%%%%%%%%%%%%%%%
\begin{figure}
\epsfysize 2.5in
%\center{\epsfbox{Fig3.eps}}
\caption{Relative densities of tilted (left panel) and non-tilted (right panel) disks .  Both panels show orbit 220, frame 0 (220.0) of 200 possible frames per orbit.  The 20$^{o}$ tilted disk shows a green higher density ring near where the gas stream strikes the disk face.  The secondary is not shown but the gas stream is shown leaving the inner Lagrange point.  In the figure, relative density is scaled orange/red=0.6\%, yellow/red=4\%, green/red=8\%, white/red=18\%, blue/red=29\%, violet/red=83\%.}
\label{Figure 3.}
\end{figure}
%%%%%%%%%%%%%%%%%%%%%%%%%%%%%%%%%%%

\clearpage

%%FIGURE 4 %%%%%%%%%%%%%%%%%%%%%%%%
\begin{figure}
\epsfysize 2.5in
%\center{\epsfbox{Fig4.eps}}
\caption{Negative superhump period excess as a function of orbital period.  Observational data are shown as stars, X's, squares, and diamonds.  IPs are squares.  AM CVn is a diamond.  Systems that only show negative superhumps in their lightcurves are X's, and systems that show both negative and positive superhumps in their light curves are stars.  Numerical data are shown as triangles.  The short  dash line represents a disk that rigidly rotates at the same rate as the primary.  The solid line represents a disk that rotates differentially with a superimposed ring on the inner edge of the disk.  The dash-dot line represents a disk that rotates differentially.  The long dash line represents a rotating ring.  Data are from Tables 1 and 2.}
\label{Figure 4.}
\end{figure}
%%%%%%%%%%%%%%%%%%%%%%%%%%%%%%%%%%%

\clearpage

%%FIGURE 5 %%%%%%%%%%%%%%%%%%%%%%%%
\begin{figure}
\epsfysize 2.5in
%\center{\epsfbox{Fig5.eps}}
\caption{Retrograde precession of a disk fully tilted 5$^{o}$ around the line of nodes, out of the orbital plane, showing that a fully tilted, precessing disk can look like a warped disk.}
\label{Figure 5.}
\end{figure}
%%%%%%%%%%%%%%%%%%%%%%%%%%%%%%%%%%%

\clearpage

%%%%%%%%%%%%%%%%%TABLE 1%%%%%%%%%%
\begin{table*}
 \centering
 \begin{minipage}{200mm}
  \caption{Negative Superhump Simulation Data}
  \begin{tabular}{@{}lllllll@{}}
  		&\multicolumn{2}{c}{Frequency ($P_{orb}^{-1}$)}& & \multicolumn{2}{c}{Period (hr)}&\\
   		\cline{2-3} 
		\cline{5-6}
q & $2\nu_{-}$ & $\sigma_{2\nu_{-}}$ & $P_{orb}$ (hr) & $P_{-}$ & $\sigma_{P_{-}}$  & $\epsilon_{n}$ \\
 \hline
0.35   & 2.098 & 0.004 & 3.540 & 3.374 & 0.006 & 0.0467 \\
0.375  & 2.105 & 0.004 & 3.698 & 3.514 & 0.006 & 0.0498 \\
0.4    & 2.104 & 0.004 & 3.852 & 3.662 & 0.006 & 0.0494 \\
0.45   & 2.110 & 0.004 & 4.150 & 3.934 & 0.007 & 0.0521 \\
0.5    & 2.115 & 0.003 & 4.436 & 4.195 & 0.007 & 0.0544 \\
0.55   & 2.119 & 0.003 & 4.712 & 4.447 & 0.007 & 0.0562 \\
\hline
\end{tabular}
\end{minipage}
\end{table*}
%%%%%%%%%%%%%%%%%%%%%%%%%%%%%%%%%%%%%%%%%%%%%%%%%%%%%

\clearpage

%%%%%%%%%%%%%%%%%%%%%%%%%%%%%%%%%%%%%%%%%%%%%%%%%Table2b
%\begin{table*}
%setlongtables
%\begin{longtable}
\begin{sidewaystable}
 \centering
 \begin{minipage}{180mm}
  \caption{Positive and Negative Superhump Observational Data, Calculated Precessional Periods, and Period Excesses}
  \begin{tabular}{@{}llllllllll@{}}
   System & Type &q  & $P_{orb}(d)$ & $P_{p_{a}}(d)$ & $P_{p_{n}}(d)$  &  $P_{a}(d)$  &  $P_{n}(d)$  & $\epsilon_{a}$ & $\epsilon_{n}$ \\
 \hline
V592 Cas$^{a}$   & 1 &$0.19^{b,+0.1}_{-0.09}$ & 0.115063(1) & 1.950 & 4.111 & 0.12228(1) & 0.11193(5) & 0.0625(5) & 0.027\\
                                 &        &  $0.248^{c}$&              &             &            &           &        &            & \\
TT Ari$^{d, e}$ & 2 &$0.25^{f}$ & 0.13755040(17) & 1.762 & 3.930 & 0.1492(1) & 0.1329 & 0.0847(7) & 0.034\\
                                 &        &  $0.315^{c}$   &             &              &            &           &        &            &\\
 		              &        &  $0.19\pm0.04^{e}$   &             &              &            &           &        &            &\\
V603 Aql$^{g,h}$ & 3 &$0.24\pm0.05^{i}$& 0.1380(5) & 2.587 & 4.409 & 0.1460(7) & 0.1340 & 0.0572(51) & 0.030  \\
                                &        &  $0.23^{c}$   &             &              &            &           &        &            &\\
V503 Cyg$^{j}$ & 4 & $0.183^{c}$ & 0.07771(28) & 1.890 & 2.921 & 0.08104(7) & 0.75694 & 0.0430(27) & 0.026 \\
V1159 Ori$^{c,k,l}$ & 5 & $0.142^{c}$& 0.0621801(13) & 2.001 & 0.7236 & 0.06417(7) & 0.05743(14) & 0.0320(11) & 0.080 \\
AM CVn$^{m, n}$ & 3 & $0.18\pm0.01^{o}$ & 0.011906(1) & 0.5627 & 0.6948 & 0.012166(1) & 0.011706 & 0.0218(1) & 0.017 \\
TV Col$^{p, q}$ & 3,6 & $\sim0.33^{p}$ & 0.229167 & 1.805 & 3.973 & 0.2625 & 0.21667 & 0.15 & 0.0545 \\
CN Ori$^{r,s}$ & 6 & & 0.163199(7) &  & 0.1421 &  & 0.1595 &  & 0.022 \\
ER UMa$^{l, t}$ & 5 & $0.140^{c}$ & 0.06366(3) & 0.4180 & 1.255 & 0.0654(5) & 0.0589(7) & 0.0314(11) & 0.074 \\
V751 Cyg$^{u}$ & 2 & & 0.144464(1) &  & 3.806 &  & 0.1394(1) &  & 0.0353(2) \\
V442 Oph$^{v}$ & 8 & $0.79\pm0.27^{w}$ & 0.12433 &  & 4.420 &  & 0.12090(8) &  & 0.027 \\
V1974 Cyg$^{x, y}$ & 3 & $0.24\pm0.04^{x}$ & 0.08126(1) & 1.805 & 2.990 & 0.08509(8) & 0.07911 & 0.0471(10) & 0.027 \\
                               &        &  $0.197^{c}$   &             &              &            &           &        &            &\\
AH Men$^{y, z}$ & 8 & $0.326^{c}$ & 0.12721(6) & 1.561 & 4.306 & 0.1385(2) & 0.12356 & 0.0887(16) & 0.029 \\
DW UMa$^{c,v}$ & 8 &$0.33^{aa}$ &0.13661& 2.2573 & 5.038 & 0.14541 & 0.1330 & 0.0644 (20)& 0.026 \\
                               &        &  $0.255^{c}$   &             &              &            &           &        &            &\\
                               &        &  $>0.24^{bb}$   &             &              &            &           &        &            &\\
PX And$^{c,y,cc}$ & 8 & $0.329(11)^{c}$&0.146353(1)&1.776&4.228&0.1595(2)&0.1415&0.0898 &0.0331 \\
BH Lyn$^{c,dd}$& 8 & $0.45^{ee,+0.15}_{-0.10}$& 0.15575(1) & 2.128 & 3.413 & 0.16805 & 0.1490 & 0.079 & 0.0433 \\
                               &        &  $0.41\pm0.26^{y}$ &             &              &            &           &        &            &\\
                               &        &  $0.301(15)^{c}$     &             &              &            &           &        &            &\\
RX J1643$^{v}$ & 8 & & 0.120560(14) & & 3.917 & & 0.11696(8) & & 0.032(2) \\
RR Cha$^{ff}$ & 6 & & 0.1401 & 4.705&5.049 & 0.1444 & 0.1363 & 0.031 & 0.027 \\
\hline
\end{tabular}
\end{minipage}
\end{sidewaystable}
%\end{longtable}

 \begin{sidewaystable}
 \centering
\begin{minipage}{180mm}
  \caption{Positive and Negative Superhump Observational Data, Calculated Precessional Periods, and Period Excesses (Continued from Table 2)} 
 \begin{tabular}{@{}llllllllll@{}}
   System & Type &q  & $P_{orb}(d)$ & $P_{p_{a}}(d)$ & $P_{p_{n}}(d)$  &  $P_{a}(d)$  &  $P_{n}(d)$  & $\epsilon_{a}$ & $\epsilon_{n}$ \\
 \hline
AT Cnc$^{gg}$  & 3,7 &0.32 - 1.04 & 0.2011(6) & & 5.051    & &0.1934(8)     & & 0.0179(10)  \\ 
                               &       & &              &  & 11.03           &  &  0.1975(8)          & &  0.0383(10)     \\
IR Gem$^{hh, ii}$ & 4 & $0.154^{c}$ & 0.06840 & 1.390& 2.105& 0.07194& 0.0663& 0.052& 0.031\\
TX Col$^{jj}$ & 6 & $>0.33$ & 0.2375 & 1.204& 1.669&0.2958&0.2083&0.25&0.1229\\
SDSS J0407$^{kk}$& 4? & &0.17017(3)& &6.727& & 0.166(1)& & 0.024(1)\\
SDSS J2100$^{ll}$  & 4& &$\sim$0.0833&1.767&4.083&0.0875&0.0817 & 0.050 & 0.020\\
CAL 86$^{mm}$ & 4 &                               &0.06613 & & 1.403 & &0.06313    & &0.045\\
KR Aur$^{nn}$    &  2 &  $0.60\pm^{oo}$ & 0.1628  & & 4.489 & & 0.1571(2) & &0.0350(2)\\
HS 1813+6122$^{pp}$& 8 & & 0.1479 & & 3.166 & & 0.1413 &  & 0.0446 \\
V2574 Oph$^{qq}$ & 9 &  & 0.14773 & & 3.429 & & 0.14164 & & 0.412 \\
RX 1643+34$^{v}$ & 8  &  & 0.12050(14) &  & 3.917 & & 0.11696(8) & & 0.03000(8) \\
AY Psc$^{rr}$          &  1   &   0.45$^{ss,0.15}_{0.20}$  &0.21732(9)  &   &  & &   0.2057(1) &                       & 0.0535\\ 
                                    &                        &  &                          &   &           & & 0.2063(1)    &  &  0.0506     \\
                                    &                        &   &                         &   &           & & 0.2072(1)    &  & 0.0466      \\
BF Ara $^{tt}$         &  1  & $\sim$0.21&0.084176(21)& & & 0.08797(1)&0.082159(4)&0.0451(3) & 0.0244(2) \\
V1193 Ori$^{uu}$ &  3      & & 0.1430 &  &  &  & 0.1362 &  & 0.0476\\

\hline
\end{tabular}
$^{1}$NL, $^{2}$VY Scl/NL, $^{3}$PS/NL, $^{4}$SU UMa/DN, $^{5}$ER UMa/DN, $^{6}$IP/NL, $^{7}$Z Cam,  $^{8}$SW Sex/NL, $^{9}$Nova, $?$classification uncertain \\
$^{a}$Taylor et al. (1998), $^{b}$Huber et al. (1998), $^{c}$Patterson et al. (2005), $^{d}$Kraicheva et al. (1999), $^{e}$Wu et al. (2002), $^{f}$Warner (1995), $^{g}$Borczyk et al. (2003), $^{h}$Patterson et al. (1997), $^{i}$Arenas (2000), $^{j}$Harvey et al. (1995),  $^{k}$Patterson et al. (1995), $^{l}$Thorstensen et al. (1997),  $^{m}$Skillman et al. (1999), $^{n}$Nelemans et al. (2001), $^{o}$Roelofs et al. (2006) $^{p}$Retter et al. (2003), $^{q}$Hellier \& Buckley (1993), $^{r}$Schoembs (1982), $^{s}$Barrera \& Vogt (1989), $^{t}$Gao et al. (1999), $^{u}$Patterson et al. (2001), $^{v}$RX J1643.7+3402 - Patterson et al. (2002), $^{w}$Shafter \& Szkody (1983), $^{x}$Retter et al. (1997), $^{y}$Patterson (1998), $^{z}$Rodriguez-Gil et al. (2007b), $^{aa}$Biro (2000), $^{bb}$Araujo-Betancor et al. (2003), $^{cc}$Boffin et al. (2003),  $^{dd}$Dhillon et al. (1992), $^{ee}$Hoard \& Szkody (1997), $^{ff}$ Woudt \& Warner (2002), $^{gg}$Kozhevnikov (2004), $^{hh}$Li et al. (2004), $^{ii}$Fu et al. (2004), $^{jj}$Retter et al. (2005), $^{kk}$SDSS J040714.78-064425.1 - Ak et al. (2005), $^{ll}$SDSS J210014.12+004446.0 - Tramposch et al. (2005), $^{mm}$Woudt et al. (2005), $^{nn}$Kozhevnikov (2007), $^{oo}$Kato et al. (2002), $^{pp}$Rodriguez-Gil et al. (2007), $^{qq}$Kang et al. (2006), $^{rr}$Gulsecen et al. (2009), $^{ss}$Szkody \& Howell (1993), $^{tt}$Olech et al. (2007), $^{uu}$Ak et al. (2005b) \\
\end{minipage}
 \end{sidewaystable}
%\end{table*}
%%%%%%%%%%%%%%%%%%%%%%%%%%%%%%%%%%%%%%%%%%%%%%%%


\begin{thebibliography}{99}
\bibitem{b0} Ak T., Retter A., Liu A., \& Esenoglu H.H., 2005, PASA, 22, 105
\bibitem{b1} Ak T., Retter A., Liu A., \& Esenoglu H.H., 2005b, NewA, 11, 147
\bibitem{b2} Araugo-Betancor A., Knigge C., Long K.S., Hoard D.W., Szkody P., Rodgers B., Krisciunas K., Dhillon V.S., hynes R.I., Patterson J., \& Kemp J., 2003, ApJ, 583, 437 
\bibitem{b3} Arenas J., Catalan M.S., Augusteijn T., \& Retter A., 2000, MNRAS, 311, 135
\bibitem{b4} Bardeen J.M. \& Petterson J.A., 1975, ApJ, 195, L65
\bibitem{b5} Barrett P., O'Donoghue D., \& Warner B., 1988, MNRAS, 233, 759
\bibitem{b6} Barrera L.H. \& Vogt N., 1989, A\&A, 220, 99
\bibitem{b7} Biro I.B., 2000, A\&A, 364, 573 
\bibitem{b8} Bisikalo D.V., Boyarchuk A.A., Kaygorodov P.V., Kuznetsov O.A., \& Matsuda T., 2004, ARep, 48, 449
\bibitem{b9} Boffin H.M.J., Stanishev V., Kraicheva Z., \& Genkov V., 2003, ASPC, 292, 297
\bibitem{b10} Bonnet-Bidaud J.M., Motch C., \& Mouchet M., 1985, 143, 313
\bibitem{b11} Borczyk W., Schwarzenberg-Czerny A., Szkody P., 2003, A\&A, 405, 663
\bibitem{b12} Boyton P.E., Crosa L.M., Deeter J.E., 1980, ApJ, 237, 169
\bibitem{b13} Caproni A. \& Abraham Z., 2002, RMxAC, 14, 74
\bibitem{b14} Danby J.M.A., 1962, in Fundamentals of Celestial Mechanics (New York: MacMillan Company)
\bibitem{b15} Dhillon V.S., Jones D.H.P., Marsh T. R., \& Smith R.C., 1992, MNRAS, 258, 225
\bibitem{b16} Eggleton P.P., 1983, ApJ, 268, 368
\bibitem{b17} Foulkes S.B., Haswell C.A., \& Murray J.R., 2006, MNRAS, 366, 1399
\bibitem{b18} Fu H., Li Z.-Y., Leung K.-C., Zhang Z.-S., Li Z.-L., \& Gaskell C.M., 2004, ChJAA, 4, 88 
\bibitem{b19} Gao W., Li Z., Wu X., Zhang Z., \& Li Y., 1999, ApJ, 527, 55
\bibitem{b20} Gulsecen H., Retter A., Liu A., Esenoglu H., 2009, NewA, 14, 330
\bibitem{b21} Harvey D., Skillman D.R., Patterson J., \& Ringwald F.A., 1995, PASP, 107, 551 
\bibitem{b22} Hellier \& Buckley D.A.H., 1993, MNRAS, 265, 766
\bibitem{b23} Hessman F.V. \& Hopp U., 1990, A\&A, 228, 387
\bibitem{b24} Hoard D.W. \& Szkody P., 1997, ApJ, 481, 433
\bibitem{b25} Huber M.E. Howell S.B., Ciardi D.R., \& Fried R., 1998, PASP, 110, 784
\bibitem{b26} Iping R.C. \& Petterson J.A., 1990, A\&A, 239,221
\bibitem{b27} Kang T.W., Retter A., Liu A., \& Richards M., 2006, AJ, 131, 1687
\bibitem{b28} Kato T., Ishioka R., \& Uemura M., 2002, PASJ, 54, 1033
\bibitem{b29} Katz J.I., 1973, Nature Phys. Sci., 246, 87
\bibitem{b30} Katz J.K., Anderson S.F., Margoon B., \& Grandi S.A., 1982, ApJ, 260, 780
\bibitem{b31} Kondo Y., Wolff C.L., \& van Flandern T.C., 1983, ApJ, 273, 716
\bibitem{b32} Kozhevnikov V.P., 2004, A\&A, 419, 1035
\bibitem{b33} Kozhevnikov V.P., 2007, MNRAS, 378, 955
\bibitem{b34} Kraicheva Z., Stanishev V., Genkov V., \& Iliev L., 1999 A\&A, 351, 607
\bibitem{b35} Kumar S., 1986, MNRAS, 223, 225
\bibitem{b36} Kumar S., 1989, in Theory of Accretion Disks, eds. F. Meyer, W.J. Duschl, J. Frank, \& E. Meyer-Hofmeister, Kluwer, Dordrecht, p. 297
\bibitem{b37} Kumar S. \& Pringle J.E., 1985, MNRAS, 213, 435
\bibitem{b38} Larwood J., 1997, MNRAS, 290, 490
\bibitem{b39} Larwood J., 1998, MNRAS, 299, L32
\bibitem{b40} Larwood J., Nelson R.P., Papaloizou J.C.B., \& Terquem C., 1996, MNRAS, 282, 597
\bibitem{b41} Larwood J. \& Papaloizou J.C.B., 1997, MNRAS, 285, 288
\bibitem{b42} Lasota J.-P., 2001, New Astronomy Review, 45, 449
\bibitem{b43} Lattanzio J.C., Monaghan J.J., Pongracic H., \& Schwarz M.P., 1986, J. Sci. Stat. Comput. 7, 591
\bibitem{b44} Li Z., Leung K.-C., \& Gaskell C.M., 2004, RMxAC, 21, 259
\bibitem{b45} Livio M., 1999, PhR, 311, 225
\bibitem{b46} Lubow S.H. \& Shu F.H., 1975, ApJ, 198, 383
\bibitem{b47} Lubow S.H., 1991a, ApJ, 381, 259
\bibitem{b48} Lubow S.H., 1991b, ApJ, 381, 268
\bibitem{b49} Lubow S.H., 1992, ApJ, 398, 525
\bibitem{b50} Lubow S.H. \& Pringle J.E., 1993, ApJ, 409, 360
\bibitem{b51} Meyer F. \& Meyer-Hofmeister E., 1981, A\&A, 104, L10
\bibitem{b52} Montgomery M.M., 2004, Ph.D Thesis, Florida Institute of Technology
\bibitem{b53} Montgomery M.M., 2009, MNRAS, 394, 1897
\bibitem{b54} Murray J.R., 1998, MNRAS, 300, 561
\bibitem{b55} Murray J.R. \& Armitage P.J., 1998, MNRAS, 300, 561
\bibitem{b56} Murray J.R., Truss M.R., \& Wynn G.A., 2002, ASPC, 261, 416 
\bibitem{b57} Nelemans G., Steeghs D., \& Groot P.J., 2001, MNRAS, 326, 621
\bibitem{b58} Nelson R.P. \& Papaloizou J.C.B., 1999, MNRAS, 309, 929
\bibitem{b59} Olech A., Rutkowski A., \& Schwarzenberg-Czerny A., 2007, AcA, 57, 331
\bibitem{b60} Osaki Y., 1985, A\&A, 144, 369
\bibitem{b61} Paczynski B., 1977, ApJ, 216, 822
\bibitem{b62} Papaloizou J.C.B. \& Terquem C., 1995, MNRAS, 274, 987
\bibitem{b63} Patterson J., 1998, PASP, 110, 1132
\bibitem{b64} Patterson J., Thomas G., Skillman D.R., \& Diaz M., 1993, ApJS, 86, 235
\bibitem{b65} Patterson J., Jablonski F., Koen C., O'Donoghue D., \& Skillman D.R., 1995, PASP, 107, 1183
\bibitem{b66} Patterson J., Kemp J., Saad J., Skillman D., Harvey D., Fried R., Thorstensen .R., Ashley R., 1997, PASP, 109, 468
\bibitem{b67} Patterson J., Kemp J., Richman H.R., Skillman D.R., Vanmunster T., Jensen L., Buckley D.A.H., O'Donoghue D., \& Kramer R., 1998, PASP, 110, 415
\bibitem{b68} Patterson J., Thorstensen J.R., Fried R., Skillman D.R., Cook L.M., \& Jensen L., 2001, PASP, 113, 72
\bibitem{b69} Patterson J., Fenton W.H., Thorstensen J.R., Harvey D.A., Skillman D.R., Fried R.E., Monard B., O'Donoghue D., Beshore E., Martin B., Niarchos P., VanMunster T., Foote J., Bolt G., Rea R., Cook L., Butterworth N., \& Wood M., 2002, PASP, 114, 1364
\bibitem{b70} Patterson J., Kemp J., Harvey D.A., Fried R.E., Rea R., Monard B., Cook L., Skillman D.R., Vanmunster T., Bolt G., Armstrong E., McCormick J., Krajci T., Jensen L., Gunn J., Butterworth N., Foote J., Bos M., Masi G., \& Warhurst P., 2005, PASP, 117, 1204
\bibitem{b71} Petterson J., 1977, ApJ, 216, 827
\bibitem{b72} Pringle J.E., 1996, MNRAS, 281, 357
\bibitem{b73} Pringle J.E., 1997, MNRAS, 292, 136 
\bibitem{b74} Retter A., Leibowitz E.M., \& Ofek E.O., 1997, MNRAS, 286, 745
\bibitem{b75} Retter A., Hellier C., Augusteijn T., Naylor T., Bedding T.R., Bembrick C., McCormick J., \& Velthuis F., 2003, MNRAS, 340, 679
\bibitem{b76} Retter A., Liu A., \& Bos M., 2005, ASSL, 332, 251
\bibitem{b77} Roberts W.J., 1974, ApJ, 187, 575
\bibitem{b78} Rodriguez-Gil P. (and 17 others), 2007, MNRAS, 377, 1747
\bibitem{b79} Rodriguez-Gil P., Schmidtobreick L., \& Gansicke B.T., 2007b, MNRAS, 374, 1359
\bibitem{b80} Roelofs G.H.A., Groot P.J., Nelemans G., Marsh T.R., \& Steeghs D., 2006, MNRAS, 371, 1231
\bibitem{b81} Romero G.E., Chajet L., Abraham Z., \& Fan J.H., 2000, A\&A, 360, 57
\bibitem{b82} Schoembs R., 1982, A\&A, 115, 190
\bibitem{b83} Shafter A.W. \& Szkody P., 1983, PASP, 95, 509
\bibitem{b84} Shakura N.I. \& Sunyaev R.A., 1973, A\&A, 24, 337
\bibitem{b85} Skillman D.R., Patterson J., Kemp J., Harvey D.A., Fried R.E., Retter A., Lipkin Y., Vanmunster T., 1999, PASP, 111, 1281
\bibitem{b86} Smak J., 1999, Acta Astronomica, 49, 391
\bibitem{b87} Smith A.J., Haswell C.A., Murray J.R., Truss M.R., \& Foulkes S.B., 2007, MNRAS, 378, 785
\bibitem{b88} Smith D.A. \& Dhillon V.S., 1998, MNRAS, 301, 767
\bibitem{b89} Stacey F.D., 1977, Physics of the Earth, Second Edition (New York: John Wiley \& Sons)
\bibitem{b90} Szkody P. \& Howell S.B., 1993, ApJ, 403, 743
\bibitem{b91} Taylor C.J., Thorstensen J.R., Patterson J., Fried R.E., Vanmunster T., Harvey D.A., Skillman D.R., Jensen L., \& Shugarov S., 1998, PASP, 110, 1148
\bibitem{b92} Thorstensen J.R., Taylor C.J., Becker C.M., \& Remillard R.A., 1997, PASP, 109, 477
\bibitem{b93} Terquem C., Papaloizou J.C.B., \& Nelson R.P., 1999, Astrophysical disks ASP Conference Series, Vol. 160, 71
\bibitem{b94} Terquem C. \& Papaloizou J.C.B., 2000, A\&A, 360, 1031
\bibitem{b95} Tramposch J., Homer L., Szkody P., Henden A., Silvestri N.M., Yirak K., Fraser O.J., \& Brinkmann J., 2005, PASP, 117, 262
\bibitem{b96} Warner B., 1995, Cataclysmic Variable Stars (New York: Cambridge U. Press)
\bibitem{b97} Warner B., 2003, Cataclysmic Variable Stars (New York: Cambridge U. Press)
\bibitem{b98} Wijers R.A.M.J. \& Pringle J.E., 1999, MNRAS, 308, 207
\bibitem{b99} Wood M.A., Montgomery M.M., \& Simpson J.C., 2000, ApJL, 535, L39
\bibitem{b100} Woudt P.A. \& Warner B., 2002, MNRAS, 335, 44
\bibitem{b101} Woudt P.A., Warner B., \& Spark M., 2005, 364, 107
\bibitem{b102} Wu X., Li Z., Ding Y., Zhang Z., \& Li Z., 2002, ApJ, 569, 418
\end{thebibliography}
\end{document}